\documentclass[prd,showpacs,showkeys,twocolumn,floatfix]{revtex4} 
\usepackage[]{epsfig,dcolumn}
\begin{document}

\title{ A Study of the $\eta \pi^{0}$ Spectrum and Search for a $J^{PC} = 1^{-+}$ Exotic Meson}
\author{A.~R.~Dzierba}
\author{J.~Gunter}
\author{S.~Ichiriu}
\author{R.~Lindenbusch}
\author{E.~Scott}
\author{P.~Smith}
\author{M.~R.~Shepherd}
\author{S.~Teige}
\affiliation{Department of Physics, Indiana University, Bloomington, IN 47405}
\author{M. ~Swat}
\author{A.~P.~Szczepaniak}
\affiliation{Nuclear Theory Center, Indiana University, Bloomington, IN 47405}
\author{S.~P.~Denisov}
\author{A.~V.~Popov}
\author{D.~I.~Ryabchikov}
\affiliation{Institute for High Energy Physics, Protvino, Russian Federation 142284}
\author{L.~I.~Sarycheva}
\affiliation{Institute for Nuclear Physics, Moscow, Russian Federation 119899}
\author{J.~Napolitano}
\affiliation{Department of Physics, Rensselaer Polytechnic Institute, Troy, NY 12180}
\date{\today}

\begin{abstract}
A partial wave analysis (PWA) of the of the $\eta \pi ^0$ system 
(where $\eta \to \gamma \gamma$) produced in the
 charge exchange reaction $\pi ^-p\to \eta \pi ^0n$ at an incident momentum
 of 18~GeV$/c$\ is presented as a function of ${\eta \pi ^0}$ invariant
 mass, $m_{\eta\pi^0}$, and momentum transfer squared, $t_{\pi^{-}\to\eta\pi}$,
from the incident $\pi^-$ to the outgoing ${\eta\pi ^0}$ system. 
 $S$, $P$ and $D$ waves were included in the 
PWA.  The 
$a_0(980)$ and $a_2(1320)$ states are clearly observed in  the overall
${\eta\pi ^0}$ effective mass distribution as well as in the amplitudes
associated with $S$ wave and $D$ waves respectively after partial
wave decomposition.   The observed distributions in moments
(averages of spherical harmonics) were 
compared to the results from the PWA and the two are consistent.  The
distribution in  $t_{\pi^{-}\to\eta\pi}$ for individual $D$ waves associated
with natural and unnatural parity exchange in the $t$-channel
  are consistent with Regge phenomenology. Of particular interest in this
study is the $P$ wave since this leads to an exotic $J^{PC}=1^{-+}$ for
the  $\eta \pi$ system.
A $P$ wave is present in the data, however attempts to describe the mass dependence
of the amplitude and phase motion with respect to the $D$ wave as a Breit-Wigner
resonance are problematic.  This has implications regarding the existence of a reported exotic
$J^{PC} = 1^{-+}$ meson decaying into $\eta \pi^0$ with a mass near 1.4~GeV$/c^2$.
\end{abstract}
\maketitle

\section{Introduction}

The observation nearly four decades ago that mesons are grouped in nonets, each characterized
by  unique values of $J^{PC}$ --
spin ($J$), parity ($P$) and charge conjugation ($C$) quantum
numbers -- led to the development of the  quark model.  Within this picture, mesons
are bound states of a quark ($q$) and antiquark ($\bar q$).  
The three light-quark flavors
($up$, $down$ and $strange$) suffice to explain the spectroscopy of most  
-- but not all -- of the lighter-mass mesons (i.e. below 3~GeV$/c^2$) 
 that do not explicitly carry heavy flavors (charm or beauty).
Early observations yielded only those $J^{PC}$ quantum numbers consistent
with a fermion-antifermion bound state.
 The $J^{PC}$ quantum numbers of a $q \bar q$ system with total quark spin, $\vec S$,
 and relative angular momentum , $\vec L$, are determined as follows: $\vec J = \vec L + \vec S$,
$P=(-1)^{L+1}$ and $C=(-1)^{L+S}$. Thus $J^{PC}$ quantum numbers such as $0^{--}$, $0^{+-}$,
$1^{-+}$ and $2^{+-}$ are not allowed and are called \emph{exotic} in this context. 

Our
understanding of how quarks form mesons has evolved within quantum chromodynamics (QCD) and
we now expect a richer spectrum of mesons that takes into account not
only the quark degrees
of freedom but also the gluonic degrees of freedom. Gluonic mesons with no quarks 
(\emph{glueballs}) are expected.  These are bound states of gluons and since the
quantum numbers of low-lying glueballs are not necessarily  exotic, they  should manifest themselves as extraneous
states that cannot  be accommodated within $\bar q q$ nonets.  Indeed there is evidence
for a glueball state \cite{CBglue}  based on overpopulation of the scalar nonet but identification is complicated
since the observed states can be a mixture of a glueball and $\bar q q$. 
Excitations of the gluonic field binding the quarks can also give rise to 
so-called \emph{hybrid} mesons that can be viewed as  bound states of a quark,
anti-quark and valence gluon ($q \bar q g$). 
 An alternative picture of hybrid mesons,
one supported by lattice QCD, is one in which a gluonic flux tube 
forms between the quark and 
anti-quark and the excitations of this flux tube
lead to so-called \emph{hybrid} mesons \cite{Isg85}. Conventional $\bar q q$ mesons arise
when the flux tube is in its ground state.
Some hybrid mesons can have a unique signature,
 exotic $J^{PC}$, and therefore the spectroscopy of 
exotic hybrid mesons is not complicated by possible mixing with
conventional    $\bar q q$ states.  
According to the flux-tube model \cite{Isg85} and lattice gauge calculations \cite{Be97}, one expects
the lightest $J^{PC} = 1^{-+}$ exotic hybrid to have a mass of about 1.9~GeV$/c^2$.
In contrast, calculations based on the MIT bag model \cite{deViron,Bar82,Jaffe} 
 place the mass of the lightest $J^{PC}=1^{-+}$ hybrid meson at about 1.4~GeV$/c^2$.

The spectroscopy of exotic
mesons provides an attractive starting point for the study of gluonic excitations.
Since flux tubes are thought to be responsible for the confinement mechanism in
QCD,  experimental information on this spectroscopy is of fundamental importance.

\subsection{Experimental situation}

After about two  decades of experimental searches there have been reports of
experimental observations
of  states with $J^{PC} = 1^{-+}$  by the Brookhaven E852 collaboration in $\pi^{-}p$
interactions at 18~GeV$/c$.  One of these has a mass of ($1593\pm 8_{-47}^{+29}$)~MeV$/c^2$\ 
and width of ($168\pm 20_{-12}^{+150}$)~MeV$/c^2$\ and decays into 
$\rho^{0} \pi^{-}$ \cite{Adams98,SUC2002} and another has a similar mass, ($1597\pm 10_{-10}^{+45}$)~MeV$/c^2$,
 but a larger width,($340\pm 40_{-50}^{+50}$)~MeV$/c^2$, and decays into
$\eta^{\prime} \pi^{-}$ \cite{Ivanov2001}. The E852 collaboration 
 also reported observation of another 
 $J^{PC} = 1^{-+}$ state with mass ($1370\pm 16_{-30}^{+50}$)~MeV$/c^2$\
and a width of ($385\pm 40_{-105}^{+65}$)~MeV$/c^2$\ decaying into $\eta \pi^{-}$
\cite{DRT97,SUC1999}. 
If an  $\eta \pi$ system
is in a $P$ wave, the resulting  $J^{PC}$ quantum number combination is exotic
($1^{-+}$).
 Critical to the identification of this state is  not only
showing the presence of a $P$ wave, but also that the resulting line shape
is consistent with a Breit-Wigner and that the
phase motion of the $P$, as determined by its interference with the dominant $D$ 
 wave cannot be due solely to the  $a_2^-(1320)$ resonance.
Soon after the E852 report, the Crystal Barrel Collaboration
reported an exotic  $J^{PC} = 1^{-+}$ state produced in 
$\bar p n \to \pi^{-} \pi^0 \eta$ obtained by stopping antiprotons in liquid
deuterium \cite{Abe98}. They reported a mass of ($1400\pm 20_{-20}^{+20}$)~MeV$/c^2$\ 
and a width of ($310\pm 50_{-30}^{+50}$)~MeV$/c^2$.  A later analysis of $\bar p p \to \pi^0 \pi^0 \eta$
supported this evidence for an exotic state \cite{Abe99}.

The first claim of an exotic meson decaying into $\eta \pi^0$ 
with a mass of  1400~MeV$/c^2$\ was made by
the GAMS collaboration in the reaction $\pi^{-} p \to \eta \pi^0  n$ \cite{Alde88}
 but a later analysis by the group \cite{Yu95} led
to ambiguous results.  The VES collaboration also presented evidence for a
P-wave contribution in $\eta \pi$ \cite{Bela93}
and at KEK a claim was made for an exotic 
$\eta \pi$ state \cite{Aoy93} as well but with a mass and width close to that of the 
$a_2(1320)$ and leakage from the dominant $D$ wave could not be excluded.

Indeed, in all the observations, the  $\eta \pi$ $P$ wave enhancements have cross sections
that are substantially smaller than the dominant $a_2(1320)$ so
 this leakage, usually due to an imperfect understanding of experimental acceptance,
is a source of concern.  For example, an observed enhancement
in the $P$ wave in the $\rho \pi$ system at low mass was reported in references \cite{Adams98,SUC2002}
as being due to leakage.  Apart from these experimental issues, the interpretation of the nature of low-mass  
$\eta \pi$ $P$ wave amplitude and phase
motion should be guided by the principle of parsimony -- less exotic interpretations must
also be considered.

This work presents a partial wave analysis (PWA) of the  $\eta \pi^{0}$ system 
 produced in  $\pi^{-}p$
interactions at an incident momentum of
18~GeV$/c$\ using the E852 apparatus.  In this analysis both the 
$\eta$ and $\pi^{0}$ are detected through their $2 \gamma$ decays.  The starting
point uses data corresponding to the reaction $\pi^{-}p \to 4 \gamma n$.  A subset
of these data correspond to $\pi^{-}p \to \pi^{0} \pi^{0} n$ and the partial wave
analysis of the $\pi^{0} \pi^{0}$ system has been  reported on in an earlier 
publication \cite{gunter2001}.
Here we present the PWA of data from $\pi^{-}p \to \eta \pi^{0} n$.  

 In contrast to the $\eta \pi^{-}$ system, charge-conjugation $C$ is a 
good quantum number for the neutral  $\eta \pi^{0}$ system.  In the E852
experiment the $\eta \pi^{-}$ and $\eta \pi^{0}$ systems have different 
systematics.  The latter relies on measurements from an electromagnetic detector alone
while the former also requires information from charged particle tracking in the forward direction.
Moreover, as will be discussed below, two resonances dominate the $\pi^{-}p \to \eta \pi^0 n$
reaction; the scalar ($S$ wave) $a_0(980)$ and the tensor ($D$ wave) $a_2(1320)$
providing benchmarks for the PWA.  In constrast, only the $a_2(1320)$ is dominant in the
$\pi^{-}p \to \eta \pi^- p$ reaction.

\subsection{Outline for this paper}

This paper is organized as follows. The experimental overview is presented 
in Section 2. Event reconstruction and data selection are described in
 Section 3. The details of the PWA formalism and 
results are given in Section 4 along with the criteria used to select the
physical solution from the set of ambiguous solutions.  In Section 5 the results of leakage 
studies are presented.  In Section 6 the distributions in moments (averages of
spherical harmonics) are compared with moments calculated from the PWA solutions.
Fits to the moment distributions are also presented.  
In Section 7 the mass and
$t$ dependence of the PWA solutions are presented and discussed.  In particular
the $t$ distributions for the $D$ waves are described within the context of Regge phenomenology.
The conclusions are summarized in Section 8.

\section{Experimental Overview}

The E852 apparatus \cite{Te98} was built around and included the Multi-Particle 
Spectrometer (MPS) at BNL. The data used for the analysis reported in this paper were 
collected in 1994
and 1995 using a nearly pure ($>$ 95\%) beam of negative pions  of 
momentum 18.3~GeV$/c$. A
30-cm-long liquid hydrogen target was surrounded by a cylindrical drift chamber
\cite{BY97} and
an array of thallium-doped CsI crystals \cite{TA96}
arranged in a barrel, all located inside the MPS
dipole magnet. Drift chambers were used to track charged particles downstream
of the target. Two proportional wire chambers (PWC's), downstream of the target,
were used in requiring specific charged particle multiplicities in the event
trigger. A 3000-element lead glass detector (LGD) \cite{Cr97} measured the
energies and positions of photons in the forward direction. The dimensions
of the LGD matched the downstream aperture of the MPS magnet. Photons missing
the LGD were detected by the CsI array or by a lead/scintillator sandwich array
(DEA) downstream of the target and
arranged around an aperture
to allow for the passage of charged particles. 

The first level trigger required that the unscattered or elastically scattered
beam not enter an arrangement of two small beam-veto
 scintillation counters located in front
of the LGD. The next level of trigger required that there be no signal in the
DEA and no charged particles recorded in the cylindrical drift chamber surrounding
the target or in the PWC's (an all-neutral trigger). In the 1994 run all layers
of the cylindrical drift chamber were used in the trigger requirement whereas
in the 1995 run, only the outer layer was used.  A common off-line analysis
criterion required no hits in the cylindrical drift chamber. The
final trigger requirement was a minimum deposition of electromagnetic energy
in the LGD \cite{Cr97} corresponding to 12~GeV.

The LGD is central to this analysis.
The LGD was initially calibrated by moving each module into a monoenergetic
electron beam. Further calibration was performed by adjusting the calibration
constant for each module until the width of the $ \pi ^{0} $ and $ \eta  $
peaks in the $ \gamma \gamma  $ effective mass distribution was minimized.
The calibration constants were also used for a trigger processor that did
a digital calculation of energy deposited in the LGD and the effective mass
of photons striking the LGD \cite{Cr97}. A laser-based monitoring system allowed for tracking the 
gains of individual modules. 

Studies were made of various algorithms for finding clusters of energies deposited
by photons, including issues of photon-to-photon separation and position finding
resolution. These are also described in reference \cite{Cr97}.

\section{Event Reconstruction and Data Selection}

Experiment E852 took data in 1994 and 1995 with 
an approximate seven-fold increase in statistics
collected in 1995 compared to 1994.  

\begin{figure}
\centerline{\epsfig{file=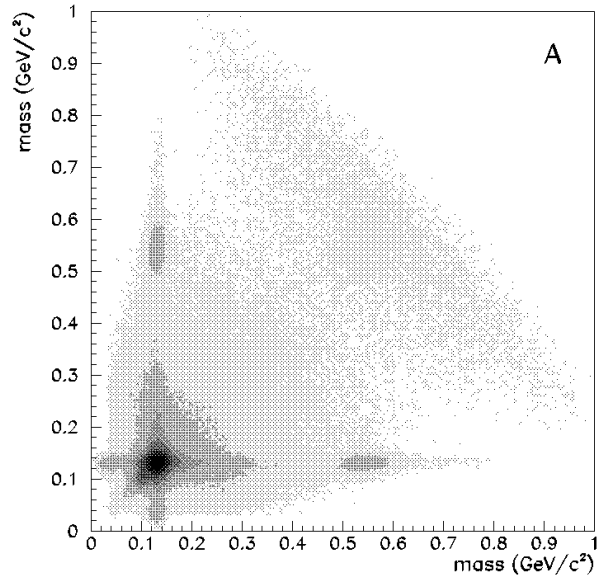,width=\columnwidth}}
\caption{ The plot of pairs of di-photon effective masses (\protect$ 
m_{ij}\protect $
vs. \protect$ m_{kl}\protect $) for all pairs of photons (\protect$ i,j,k,l\protect $) is 
dominated by the \protect$ \pi ^{0}\pi ^{0}\protect $ signal. Clear evidence
is also seen for the production of \protect$ \eta \pi ^{0}\protect $. }
\label{scatter}
\end{figure}

The combined data sets taken in 1994 and 1995 contain approximately 70 million
all-neutral triggered events. Of these events, approximately 13 million were found
to have four photons in the LGD. Figure \ref{scatter} shows a scatterplot of 
one di-photon effective mass against the other di-photon combination in the
event.  There are clear indications of the presence of $\pi^0 \pi^0$ and $\eta \pi^0$ 
events in the sample.  The single clustering at $m_{\gamma \gamma}$ = $m_{\gamma \gamma}$ =0.135~GeV$/c^2$\
is dominated by $\pi^0 \pi^0$ events and the two clusterings at $m_{\gamma \gamma}$ =0.135~GeV$/c^2$,
$m_{\gamma \gamma}$ =0.540~GeV$/c^2$\ are dominated by $\eta \pi^0$ events.

Studies were carried out comparing the 1994 and 1995 data sets for both  $\pi^0 \pi^0$ and $\eta \pi^0$
events.  In each case distributions in the decay angles of the $\pi^0 \pi^0$ (or $\eta \pi^0$)
system were studied in mass bins of 40~MeV$/c^2$\ and in four bins of $t$,
where $t$ is the square of the momentum transferred from the beam $\pi^-$ to the 
outgoing $4\gamma$ system. A Kolmogorov test
measured the probability that the two distributions (1994 and 1995 data) in each case were
consistent with having been drawn from the same parent distribution.  The conclusion of this
study is that the 1994 and 1995 data sets are statistically indistinguishable.

 Events consistent
with the production of two $ \pi ^{0} $'s dominate the sample.  The partial
wave analysis of the  $\pi^0 \pi^0$ events is presented in an unpublished
thesis of J. Gunter \cite{jgth} and in a previous paper \cite{gunter2001}.  The identification 
of events from the reaction $\pi^- p \to \eta \pi^0 n$ is described in the 
unpublished thesis of R. Lindenbusch \cite{roblin}.

 The sample
of 45,000 $ \pi ^{-}p\rightarrow \eta\pi ^{0}n $ events was selected
from the 13 million four photon events by  imposing various analysis criteria.  
It was required that no charged
particles were registered in the MPS drift chambers or the cylindrical drift chamber 
surrounding the liquid hydrogen target. This cut was performed at the pre-reconstruction
level and was based on total hit multiplicity.
Any event with a photon within  8 cm  of the center of the beam 
hole or the
outer edge of the LGD was removed.

Monte Carlo studies were carried out to determine the $ \chi ^{2} $ criteria needed
to select $\eta \pi^0$ events and eliminate backgrounds from other processes, especially
from $\pi^0 \pi^0$ events.  Events corresponding to $\pi^0 \pi^0$, $\eta \pi^0$, $\eta \eta$
and $\eta^{\prime} \pi^0$  were generated and passed through detector simulation, 
reconstruction, kinematic fitting and other event selection software.  
The $ \chi ^{2} $ returned from kinematic fitting to the 
$\pi p \to \eta\pi^{0}n $ 
reaction  hypothesis was required to be less than 7.8 (95\% C.L. for a
three-constraint fit).  The two-constraint $ \chi ^{2} $ for the $ \pi^{0} \pi^{0} $
hypothesis (no requirement that the missing mass be consistent with the neutron mass)
 was required to be greater than 100.  It was found that this latter cut eliminated 
true $ \pi^{0}\pi^{0} $ events that have a poor fit to the $\pi p \to \pi^{0}\pi^{0}n $
hypothesis.
 A further demand was that none   of the
other final state hypotheses considered 
($\eta \eta n$, $\eta^{\prime} \pi^0 n$) had a better $ \chi ^{2} $ than that used
to select the $\eta \pi^0 n$ hypothesis.

The final criterion was that the CsI detector registered
 less than 20 MeV, a cut which eliminated events with
low-energy $ \pi ^{0} $'s. Background studies \cite{roblin} estimated the non-$\eta$
background in the $\pi p \to \eta\pi^{0}n $.  The signal-to-noise ratio varies from
about $5.1:1$ in the $\eta\pi^{0}$ mass region 1.2 to 1.5~GeV$/c^2$\ to 
$2.5:1$ in the $\eta\pi^{0}$ mass regions below this range and above this range up
to 1.8~GeV$/c^2$.

\begin{figure}
\centerline{\epsfig{file=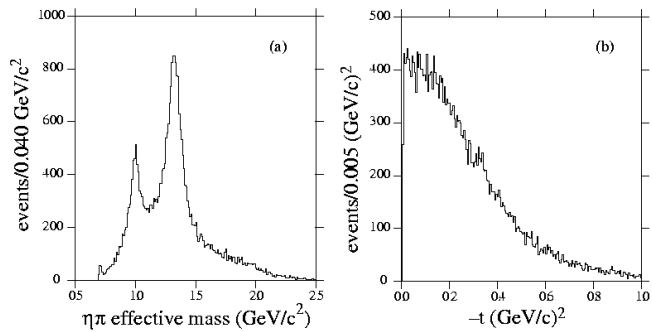,width=\columnwidth}}
\caption{ Distribution in the (a)  $\eta \pi^0$ effective mass;
(b)  momentum-transfer-squared ( $-t_{\pi^{-}\to\eta\pi}$) 
from the incoming
$\pi^-$ to the outgoing $\eta \pi^0$ system. These distributions are not corrected
for losses due to acceptance. }
\label{m_and_t}
\end{figure}

The  $\eta\pi^{0}$ effective mass distribution is shown  in
figure \ref{m_and_t}(a).  The  $a_0(980)$ and $a_2(1320)$ states, one
a scalar and the other a tensor, are clearly observed.
The presence of both states is important for this analysis since
it provides a basis, as will be discussed below, for selecting
the physical solutions from among mathematically ambiguous solutions in doing the
partial wave analysis.
This distribution in $\eta\pi^{0}$ effective mass
 is not corrected for the acceptance of the apparatus
but the PWA procedure does take into account the effects of experimental
acceptance. 

\begin{figure}
\centerline{\epsfig{file=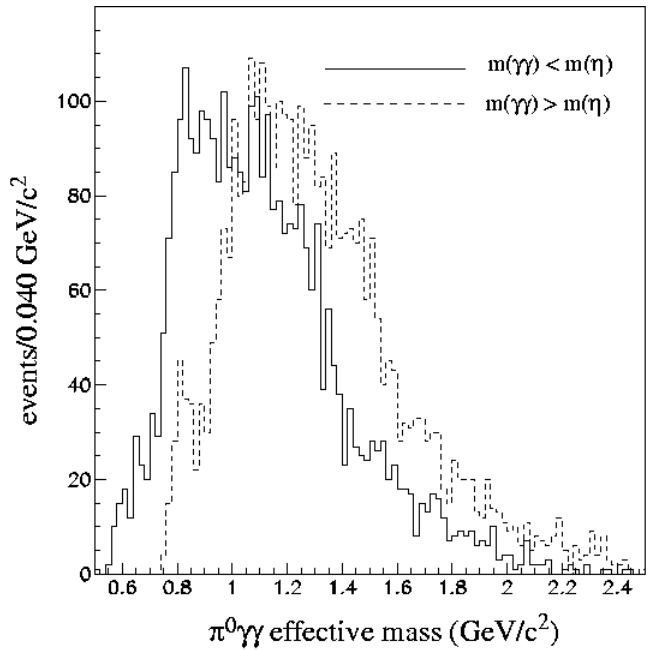,width=\columnwidth}}
\caption{ Distribution in the four photon effective mass for events with a reconstructed $\pi^0$ and the
other $\gamma \gamma$ effective mass combination falling below (solid) and
above (dashed) the $\eta$ mass region. Neither distribution shows the structure observed
for $\eta \pi^0$ events. }
\label{sidebands}
\end{figure}

The distribution in  $t_{\pi^{-}\to\eta\pi}$ is shown in
figure \ref{m_and_t}(b).
Figure \ref{sidebands} shows the four-photon effective mass distribution for
events with a di-photon mass combination consistent with a $\pi^0$ and the
other di-photon mass either below or above the $\eta$ mass region for a portion
of the total four-photon sample.  The resulting $ \gamma \gamma \pi^0$ spectra
do not show the enhancements at the $a_0(980)$ and $a_2(1320)$ observed
in the $\eta \pi^0$ spectrum.  

A possible source of background in the $\eta \pi^0$ in the $a_0(980)$ mass
region are events corresponding to production of the final state $\eta \pi^0 \pi^0 n$
and in particular that subset corresponding to $f_1(1285)n$ where the 
$f_1$ decays to $a_0 \pi^0$.  There events can enter the sample under study
if one of the $\pi^0$'s escapes detection.  From a Monte Carlo study of this
sample \cite{roblin} based on a measurement of $f_1(1285)$ production in this
experiment \cite{Te98} it is estimated that at most 10\% of the events in the
$a_0(980)$ region are due to this background process.

\begin{figure}
\centerline{\epsfig{file=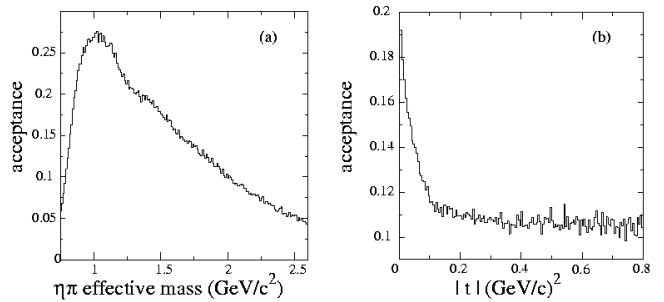,width=\columnwidth}}
\caption{Acceptance as a function of  (a)  $\eta \pi^0$ effective mass;
(b)  momentum-transfer-squared ( $-t_{\pi^{-}\to\eta\pi}$) 
from the incoming
$\pi^-$ to the outgoing $\eta \pi^0$ system.}
\label{accep_m_and_t}
\end{figure}

The acceptance in $\eta\pi^{0}$ effective mass and $t_{\pi^{-}\to\eta\pi}$
was estimated by generating Monte Carlo events corresponding to the reaction 
$ \pi ^{-}p\rightarrow \eta\pi ^{0}n $. The acceptance functions as a
function of $\eta \pi^0$ effective mass are shown in
figure \ref{accep_m_and_t}.  
The acceptance in $t_{\pi^{-}\to\eta\pi}$
increases for $-t$ near zero.  This comes about because of the all-neutral 
requirement in the trigger. The recoil neutron produced in the reaction can
interact and cause a signal in the CsI detector thereby vetoing true $\pi^0 \pi^0 n$
events but for small values of $|t|$ the neutron cannot escape the LH$_2$ target and
the event will not be vetoed.

\begin{figure}
\centerline{\epsfig{file=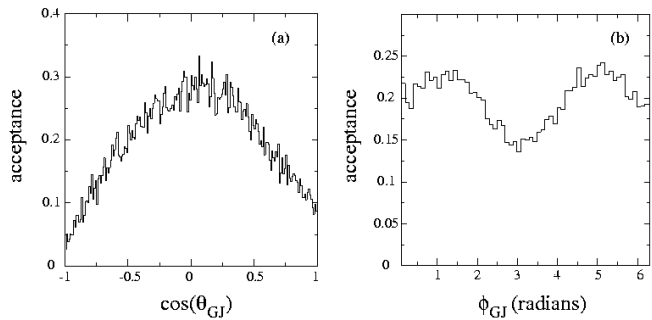,width=\columnwidth}}
\caption{  
Acceptance in Gottfried-Jackson angles for the $\eta \pi^0$ effective
mass region from 1.3 GeV$/c^2$\ to 1.4 GeV$/c^2$\ for:
(a) $cos(\theta_{GJ})$. (b) $\phi_{GJ}$. }
\label{accep_thet_and_phi}
\end{figure}

\section{Partial Wave Analysis}

Partial wave analysis is used to extract production amplitudes (partial waves)
from the observed decay angular distributions of the $\eta \pi^0$ system.  A  
process such as $\pi^-p\rightarrow \eta \pi^0 n $, dominated by $t$-channel meson exchange, 
is  conveniently analyzed in the Gottfried-Jackson
reference frame. The Gottfried-Jackson frame is defined as a right-handed coordinate
system in the center of mass of the produced $\eta \pi^0$ system with the $ z-axis $
defined by the beam particle momentum and the $ y-axis $ perpendicular
to the plane defined by the beam and recoil neutron momenta. The decay angles
($ \theta ,\phi  $) are determined by the $ \eta $
four-momentum vector. At a fixed beam momentum, an event is fully specified by $ (m_{\eta \pi },t,\theta 
,\phi ) $.   The acceptances of the E852 apparatus for the Gottfried-Jackson angles
were studied using Monte Carlo simulations.  Results for the $\eta \pi^0$ effective
mass region from 1.3~GeV$/c^2$\ to 1.4~GeV$/c^2$\ are shown in figure \ref{accep_thet_and_phi}.

The
production amplitudes and their relative phases are extracted from the
accumulated angular distributions, for bins in $ m_{\eta \pi } $ and $ t $,
using  an extended maximum likelihood fit to the 
distributions in ($\theta$,$\phi$) \cite{jgth,roblin,Chu97}.
Fits were performed for  three bins in $t_{\pi^{-}\to\eta\pi}$ (or 
simply $t$) which we label 
as low-$|t|$ for $|t| <$ 0.14~(GeV$/c$)$^2$, medium-$|t|$ for 0.14 $<|t| <$ 0.31~(GeV$/c$)$^2$,
and high-$|t|$ for $|t| >$ 0.31~(GeV$/c$)$^2$.  For each $t$ bin the PWA fits were
performed in 0.04~GeV$/c^2$-wide  $\eta \pi^0$ mass bins.

Events corresponding to the  reaction $ \pi ^{-}p\rightarrow \eta\pi ^{0}n $ were
processed through PWA software that extracts partial wave amplitudes and phase
differences between various partial waves.  This procedure takes into account 
information about the acceptance of the apparatus.  The PWA formalism used in this
analysis is model independent since the three-body final state follows from a
quasi-two body interaction.  This is in contrast to the \emph{isobar} model
needed for more complex final states, for
example, in which one makes specific assumptions about a series of sequential
decays leading to the final state under consideration.

The decay angular distributions are described by an expansion in partial waves
that are products of a production amplitude for a state of angular momentum $L$ and
associated magnetic quantum number $M$ and a decay amplitude, also for a given
$L$ and $M$.  The angular momentum $L$ is between the $\eta$ and the $\pi^0$.
The production amplitude takes into account the  spin degrees of freedom of the 
initial and final nucleons and the naturality of the exchange particle.  That is,
the assumption is made that in this peripheral process the produced 
$\eta \pi^0$ state results from a $t$-channel exchange from the nucleon
vertex to the meson production vertex.  Alternatively one can view the production
vertex as a time-reversed decay of the produced $\eta \pi^0$ state into the beam
particle ($\pi$) and the exchanged particle. 

 Parity conservation in the strong interactions implies a relationship between
the magnetic quantum number $M$ and naturality.
Following \cite{Chu97} we choose  
 a basis  characterized by the naturality of the exchanged particle.  \emph{Natural}
parity exchange obtains
 if parity and spin are related as $P = (-1)^J$ (e.g. 
 $\rho$ meson exchange) and the exchange is \emph{unnatural} if $P = (-1)^{J+1}$ 
(e.g.   $\pi$ meson exchange). 

This basis, the \emph{reflectivity} basis, is one that does not
possess the mixed symmetries of the D-functions or spherical harmonics under rotations about the $y$
axis. For $M >0$ the linear combination with positive reflectivity, corresponding to natural parity exchange,
is:
\begin{equation}
{^+}D^{L}_{M0}(\phi,\theta,0)\equiv\frac {1}{\sqrt{2}}[D^{L}_{M0}(\phi,\theta,0)-(-1)^MD^{L}_{-M0}(\phi,\theta,0)]  
\end{equation}
and the linear combination with negative reflectivity, corresponding to unnatural parity exchange, is:
\begin{equation}
{^-}D^{L}_{M0}(\phi,\theta,0)\equiv\frac {1}{\sqrt{2}}[D^{L}_{M0}(\phi,\theta,0)+(-1)^MD^{L}_{-M0}(\phi,\theta,0)]  
\end{equation}

For $M=0$ the reflectivity is negative and:

\begin{equation}
{^-}D^{L}_{00}(\phi,\theta,0) \equiv D^{L}_{00}(\phi,\theta,0)
\end{equation}

 The distribution of  events in  $m_{\eta \pi}$, $t$, $\theta$ and $\phi$ is given by 
\begin{eqnarray}
I(m_{\eta\pi},t,\theta,\phi) 
 &=& |\sum_{LM} A^+_{LM}(m_{\eta\pi},t)\; ^+D^{L}_{M0}(\theta,\phi,0) |^2\nonumber \\ 
 &+& |\sum_{LM} A^-_{LM}(m_{\eta\pi},t)\; ^-D^{L
 *}_{M0}(\theta,\phi,0) |^2, \label{intensity}
\end{eqnarray}

where the $A_{LM}$ are the production amplitudes and we  refer to the various waves as 
 $S_0 \equiv A^-_{00}$, $P_+ \equiv A^+_{11}$, $P_- \equiv A^-_{11}$, 
 $P_0 \equiv A^-_{10}$, $D_+ \equiv A^+_{21}$, $D_- \equiv A^-_{21}$, 
and  $D_0 \equiv A^-_{20}$.

 In this analysis we considered $L=0,1,2$ (the nomenclature
is $S$, $P$ and $D$ respectively).  Waves are further characterized by a subscript
$0$ if $M=0$ and $+$ or $-$ if $|M|=1$.  As noted in \cite{Chu97}, amplitudes with
$|M|>1$ are not expected to be important for the production of mesons in $\pi p$
quasi-two-body processes. This assumption is borne out by PWA
fits to these data indicating little contribution from $|M|>1$.
We present these results after the discussion of fitting the moments
distributions.
The amplitudes considered in this analysis, $S_0$, $P_0$, $P_-$, 
$D_0$, and 
$D_-$ are produced via {\it unnatural } exchange and $P_+$ and  $D_+$ 
are 
produced via {\it natural} exchange. Waves of different naturality do not 
interfere. Since the nucleon polarization is not being measured, in 
 equation ~(\ref{intensity}) the dependence on the nucleon helicity
 is implicit in the production amplitude, e.g. 
 $|D_+|^2 =  |D^{++}_+|^2 + |D^{+-}_+|^2$ with the upper labels
 referring to the target and recoil nucleon helicities and the 
 two amplitudes representing nucleon helicity flip and  non-flip.

\subsection{Ambiguities and finding the physical solution}

There are multiple discrete sets of partial wave amplitudes that can give rise
to exactly the same angular distribution \cite{Chu97}.  It can be shown that 
for the production of two pseudoscalars  with only $S$, $P$ and $D$ waves
there are up to eight ambiguous solutions.  For a given bin in $m_{\eta \pi^0}$ these
multiple solutions can be found by randomly seeding the fit 
(using sets of different initial values for the amplitudes) 
and doing multiple fits.

It is also possible to find the ambiguous solutions analytically, starting with
one of the solutions found from the fitting procedure
using the method of Barrelet zeroes \cite{Bar72}.  In this analysis the ambiguous solutions 
found from 
multiple  randomly seeded fits were found to be consistent with the
ambiguous solutions found analytically.

\begin{figure}
\centerline{\epsfig{file=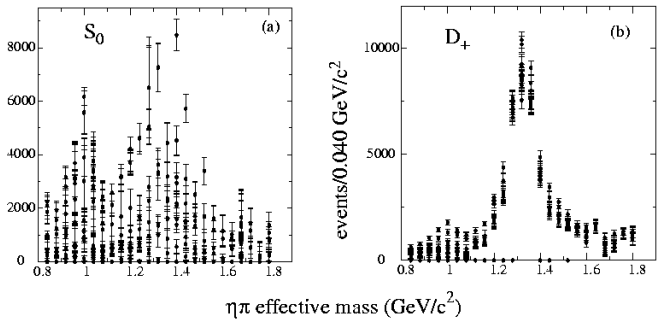,width=\columnwidth}}
\caption{PWA solution for the (a) $S_0$ and (b) $D_+$ waves as a function of 
$\eta \pi^0$ effective mass. All mathematically ambiguous solutions are shown. }
\label{s0_d+_amb}
\end{figure}

As an example, all ambiguous solutions are shown for the $S_0$ and $D_+$ waves
for the full range in $t$
 in  figure \ref{s0_d+_amb}.
The presence of the dominant $a_2(1320)$ in the $D$ wave leads to a clustering
of solutions about the expected Breit-Wigner lineshape.  The solutions for the
$S$ wave are more widely distributed.

When possible, mass-bin to mass-bin continuity along with other independent
criteria must be used to select the physical solutions among the ambiguous
solutions.  In this analysis the presence of the well-established $a_0(980)$
and $a_2(1320)$ states, as observed in the  $\eta \pi$ mass
distribution, provides a means by which to  select the physical solution. For each
$ m_{\eta \pi } $ mass bin the
solution selected as \emph{physical} is one where the $S_0$ and $D_+$ amplitudes and the 
phase difference, $\Delta \Phi(D_0 - S_0)$, all as a function of $\eta \pi$ mass, are
simultaneously  most  consistent with two interfering Breit-Wigner amplitudes with
 masses and widths for the $a_0(980)$ and $a_2(1320)$ as listed in the Review of Particle
Properties \cite{Hag02}. For example, referring to  figure \ref{s0_d+_amb},
 the $S_0$ amplitude of the physical solution
in the 1.0~GeV$/c^2$\ region should be large and its corresponding $D_+$ amplitude 
 small and in the 1.3~GeV$/c^2$\ mass region the $D_+$ amplitude should be large
and the $S_0$ amplitude should be small.
 It was found that
the requirement that the $S_0$  and $D_+$ solutions simultaneously lie
close to the expected line shapes leaves a few ambiguities  that are removed
when the criterion of proximity to the expected
value of $\Delta \Phi(D_0 - S_0)$ is imposed.  We note that although the 
$a_2(1320)$ is dominant in the $D_+$ wave, the criterion for selection involved
the phase difference between the  $D_0$ and $S_0$ waves   -- the $D_+$ and $S_0$
waves  do not interfere.
In summary, these criteria were sufficient
to   uniquely select the physical solution.

In the discussion below that follows the description of fits to experimentally 
observed moments distributions, we discuss additional quantitative measures for
discarding other ambiguous solutions as well as the effect of their inclusion on the
stability of the final results.  
In addition, as discussed below, the 
 $t$-dependences of the selected
$D$-wave solutions are well-described by Regge phenomenology  as is the
ratio ${{\left( {D_0-D_-} \right)} \mathord{\left/ 
{\vphantom {{\left( {D_0-D_-} \right)} {D_+}}} \right. \kern-\nulldelimiterspace} {D_+}}$,
 giving us confidence in the selection procedure.

\begin{figure}
\centerline{\epsfig{file=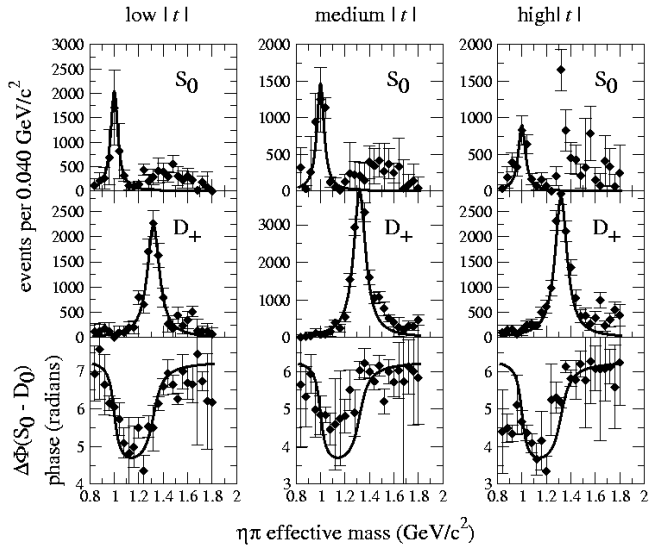,width=\columnwidth}}
\caption{Physical PWA solution for the (top) $S_0$ (middle) $D_+$ waves
and (bottom) $\Delta \Phi(S_0 - D_0)$ phase difference for
low-$|t|$, medium-$|t|$ and high-$|t|$ as a function of 
$\eta \pi^0$ effective mass. The curves are the Breit-Wigner line
shapes used as criteria for selecting the physical solution and
more details are given 
 in the text. }
\label{criteria}
\end{figure}

In figure \ref{criteria}  the selected physical solution in
each of the low-$|t|$, medium-$|t|$ and high-$|t|$ regions is shown as
a function of $\eta \pi^0$ mass for the $S_0$-wave, the $D_+$ wave and
the $\Delta \Phi (S_0 - D_0)$ phase difference. Also shown are the
corresponding line shapes and phase differences assuming two interfering
Breit-Wigner forms for the $a_0(980)$ and $a_2(1320)$ -- the basis for the
criteria used to select the physical solutions.  
We note that the intensity and phase motion of the $S_0$-wave is also consistent with
the presence of both the $a_0(980)$ and  the $a_0(1450)$. The $a_0(1450)$
resonance parameters have been extracted from a coupled channel analysis of 
final states producted in $\bar p p$ annihilations \cite{a01450}.  
This may account for
the excess 
of events in the 1.4~GeV$/c^2$\ mass region for the  $S_0$-wave intensity.  However 
because of the large statistical errors in this region we have not
been able to unambigously extract the $a_0(1450)$ resonance
 parameters.  We will return to this issue in Sections VI and
 VII.

\section{Leakage Studies}

Incomplete or incorrect understanding of the experimental acceptance
can lead to serious systematic effects in this type of analysis.
In particular, it is possible that part of a large, dominant
partial wave (for example $ D_+$) could  appear as a spurious
signal in a different partial wave. 
We refer to this phenomenon as  {\em{leakage}} and study it by 
intentionally using an incorrect model of the acceptance.

A Monte Carlo sample of pure $a_2(1320) $ events was generated
including both $ D_0 $ and $ D_+ $ waves.
This Monte Carlo sample included the effects of experimental 
acceptance. 
This sample was analyzed in two ways, in one case our best estimate
of the acceptance was used, in the second no correction for the
acceptance was made.

\begin{figure}
\centerline{\epsfig{file=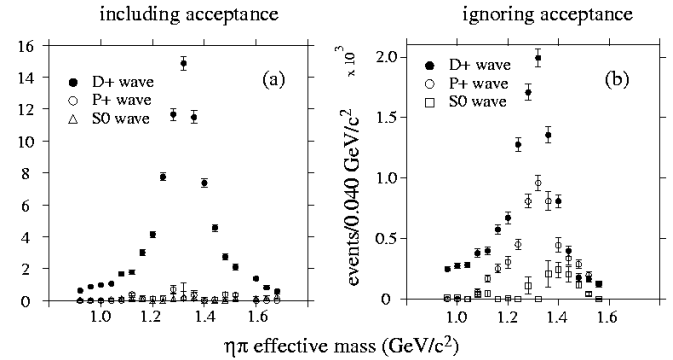,width=\columnwidth}}
\caption{Results of leakage studies.  A sample of Monte Carlo events was generated
corresponding to production of the $a_2(1320)$ populating $D$ waves.  The sample was
passed through cuts simulating detector acceptance and then through the PWA fitting
software.  In one case (a) the detector acceptance was used in the fitting and in the
other (b) the acceptance information was ignored.}
\label{leakage}
\end{figure}

The analysis using the acceptance correction should produce as
output only $ D_0 $ and $ D_+ $ waves with all others consistent
with zero. 
This was found to be the case and the resulting   $S_0$  and $ P_+ $ solutions
are compared to the $ D_+ $ intensity in figure \ref{leakage}(a). 
A similar result was obtained for all other waves.
We conclude that there are no pathological
defects in our method since we were able to reproduce correct
output from known input.

The second analysis provides an upper limit as to the 
contribution of leakage in this analysis.
The acceptance is known to be significantly different
than 100\% and a function of the kinematic variables,
see, for example, figure  \ref{accep_m_and_t}.
Approximating the acceptance as equal to 100\% independent
of  the kinematic variables simulates a situation where
our understanding of the acceptance is as incorrect as possible,
thus providing a limit on the effect of leakage.
 
It was found that ignoring the acceptance does generate spurious
signals in other partial waves. 
Results for the $ P_+ $ and $ S_0 $ partial waves are shown in
figure \ref{leakage}(b).
Crucial to the interpretation of this spurious  $ P_+ $ partial wave
is the fact that even  when ignoring corrections for
the acceptance, the $\Delta \Phi (P_+ - D_+) $  phase difference is
independent of mass.
As we will show below, the observed $\Delta \Phi (P_+ - D_+) $ phase difference
is a strong function of $\eta \pi^0$ mass.

\section{Experimental  Moments}

To check the results of PWA fitting, moments of the
spherical harmonics were calculated for each bin in $ m_{\eta \pi^0} $
for each of the three $t$ regions.
These moments are labelled as $ H(LM) $ and are defined as
\begin{equation}
H(LM) = \int I(\Omega) D^{L\ast}_{M0}(\phi,\theta,0) d\Omega.
\end{equation}
Since
\begin{equation}
H(00) = \int I(\Omega) d\Omega
\end{equation}
$ H(00)/(4\pi) $ is just the number of events.
Moments are directly measured observables and therefore not
subject to ambiguities.
Given a set of amplitudes, the expected values of the moments
can be calculated.
If the partial wave expansion is limited to $ S, P $ and $ D $
waves with $|M| \le 1$ there are 12 non-zero moments and the following
relations hold:

\begin{eqnarray}
\nonumber H(00)=\left|{S_0}\right|^2+\left|{P_0}\right|^2+\left|{P_-}\right|^2+\left|{D_0}\right|^2\\
+\left|{D_-}\right|^2+\left|{P_+}\right|^2+\left|{D_+}\right|^2
\end{eqnarray}

\begin{eqnarray}
\nonumber H(10)={2 \over \sqrt{3}}Re\{S_0P_0^*\}+{4 \over \sqrt{15}}Re\{P_0D_0^*\}\\
+{2 \over \sqrt{5}}Re\{P_-D_-^*\}+{2 \over \sqrt{5}}Re\{P_+D_+^*\}
\end{eqnarray}

\begin{eqnarray}
\nonumber H(11)={2 \over \sqrt{6}}Re\{S_0P_-^*\}+{2 \over \sqrt{10}}Re\{P_0D_-^*\}\\
-{2 \over \sqrt{30}}Re\{P_-D_0^*\}
\end{eqnarray}

\begin{eqnarray}
\nonumber H(20)={2 \over \sqrt{5}}Re\{S_0D_0^*\}+{2 \over 5}\left|{P_0}\right|^2-{1 \over 5}\left|{P_-}\right|^2\\
-{1 \over 5}\left|{P_+}\right|^2+{2 \over 7}\left|{D_0}\right|^2+{1 \over 7}\left|{D_-}\right|^2+{1 \over 7}\left|{D_+}\right|^2
\end{eqnarray}

\begin{eqnarray}
\nonumber H(21)={2 \over \sqrt{10}}Re\{S_0D_-^*\}+{2 \over 5}{\sqrt{3 \over 2}}Re\{P_0P_-^*\}\\
+{2 \over 7\sqrt{2}}Re\{D_0D_-^*\}
\end{eqnarray}

\begin{eqnarray}
\nonumber H(22)={1 \over 5}{\sqrt{3\over2}}\left|{P_-}\right|^2-{1 \over 5}{\sqrt{3\over2}}\left|{P_+}\right|^2\\
+{1 \over 7}{\sqrt{3\over2}}\left|{D_-}\right|^2-{1 \over 7}{\sqrt{3\over2}}\left|{D_+}\right|^2
\end{eqnarray}

\begin{eqnarray}
\nonumber H(30)={6\over7}{\sqrt{3\over5}}Re\{P_0D_0^*\}-{6 \over 7\sqrt{5}}Re\{P_-D_-^*\}\\
-{6 \over 7\sqrt{5}}Re\{P_+D_+^*\}
\end{eqnarray}

\begin{equation}
H(31)={{4\over7}{\sqrt{3\over5}}}Re\{P_0D_-^*\}+{{6\over7\sqrt{5}}}Re\{P_-D_0^*\}
\end{equation}

\begin{equation}
H(32)={2\over7}{\sqrt{3\over2}}Re\{P_-D_-^*\}-{2\over7}{\sqrt{3\over2}}Re\{P_+D_+^*\}
\end{equation}

\begin{equation}
H(40)={2\over7}\left|{D_0}\right|^2-{4\over21}\left|{D_-}\right|^2-{4\over21}\left|{D_+}\right|^2
\end{equation}

\begin{equation}
H(41)={2\over7}{\sqrt{5\over3}}Re\{D_0D_-^*\}
\end{equation}

\begin{equation}
H(42)={\sqrt{10}\over21}\left|{D_-}\right|^2-{\sqrt{10}\over21}\left|{D_+}\right|^2
\end{equation}

We note that  the $H(1M)$ and $H(3M)$ moments involve only terms linear
in the $P$-wave
and the $H(4M)$ moments only involve $D$-waves. The above equations 
also underscore the inherent ambiguities in this
partial wave analysis.  Solving for the amplitudes and their interferences
in terms of the non-ambiguous observable moments does not lead to a unique solution.

\begin{figure}
\centerline{\epsfig{file=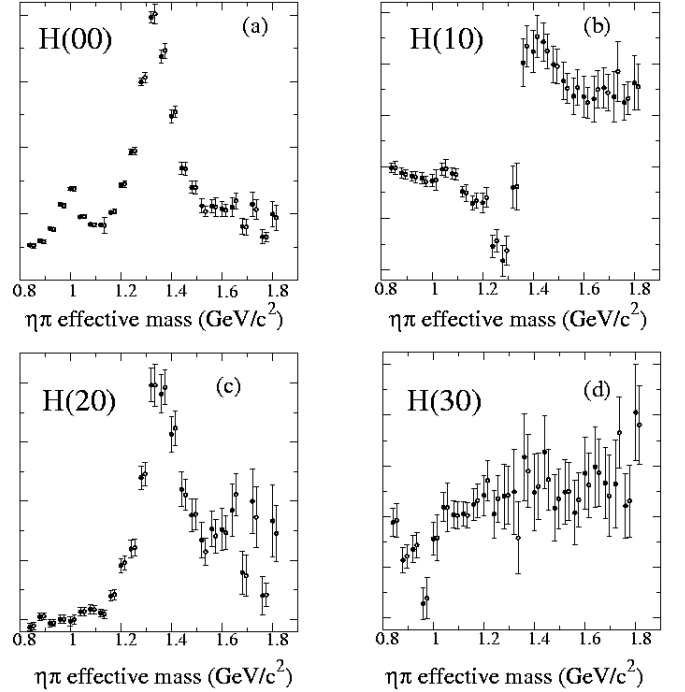,width=\columnwidth}}
\caption{Comparison between experimentally observed moment distributions
(filled circles)  and
moments calculated using PWA solutions (open circles) as a function of 
$\eta \pi^0$ effective mass for the moments: (a) $H(00)$,(b) $H(10)$, (c) $H(20)$,
and (d) $H(30)$.  The relation between moments and PWA amplitudes and phase differences
is given in the text.}
\label{comparison}
\end{figure}

In figure \ref{comparison} we show the comparison of moments computed directly
from data with those calculated from the above expressions using the
selected PWA solutions  for the 
$H(00)$,$H(10)$, $H(20)$ and $H(30)$ moments.  The agreement is excellent
and similar results were obtained for the other moments.

\subsection{Fitting the moments}

The moments as  function of $\eta \pi^0$ mass were  fitted using the
above equations and assuming Breit-Wigner resonance line shapes for the 
$a_0(980)$, the $a_2(1320)$ and a possible $P$-wave resonance.  These line
shapes specify the form of the mass dependence of the amplitudes and 
phases (interference terms) in the above equations.

  A relativistic Breit-Wigner form was used:

\begin{equation}
A_{LM}(m) =  {{A_{LM} B_L( (q R)^2 ) }\over { m^2 - m_r^2 + i m_r
\Gamma_L(m) }}
\end{equation}
where $q=\lambda(m,m_\pi,m_\eta)$ is the breakup momentum of mass $m$.
$A_{LM}$ is a complex, mass-independent normalization, and $m_r$ is mass of
the
resonance. The energy dependent width is given by
\begin{equation}
\Gamma_L(m) = \Gamma_r \left( {q\over q_r} \right)
 \left({{m_r}\over m}\right) \left( {{B_L((q R)^2 )}\over {B_L( (q_r
R)^2 )}} \right)^2
\end{equation}
with $q_r = \lambda(m_r,m_\pi,m_\eta)$ and
the barrier factors $B_L(x)$ given by
$B_0(x) = 1$, $B_1(x) = \sqrt{x/(1+x)}$, $B_2(x)=\sqrt{x^2/((x-3)^2+9
x)}$,
 where the $R$ parameterizes the range of the interaction.
In the cross section, phase space introduces
a factor of $q$ which multiplies the intensity function given in
equation~(\ref{intensity}). Finally the $S$-wave contribution was modified by
 adding  an incoherent background parametrized as a linear function in
 $m_{\eta\pi^0}$.

There are 21 parameters allowed to vary in the simultaneous fits to the distributions
of the 12 moments as a function of $\eta \pi^0$ mass.  The $S$, $P$ and $D$-wave
Breit-Wigners are each characterized by a mass and width for a total of six parameters.  Each of the
seven waves is characterized by a complex number but the overall phase for each
reflectivity is fixed accounting for 12 more parameters.  Finally the slope and intercept
of the background and the R-parameter account for the final three parameters.
  The values of these parameters
resulting from the fits are shown  in  table \ref{fits_mom}. 
 We have also attempted to include a second $S_0$-wave resonance to
account for  a possible  $a_0(1450)$ in the data. The moment fits 
were, however, not able to constrain the $a_0(1450)$ resonance
 parameters. We therefore removed the $a_0(1450)$ from the moments fits. This
  has no effect on the results and  interpretation of
 the $P$-wave fits.

\begin{table}
\begin{tabular}{lcccc}
\hline \hline
 & low-$|t|$   & medium-$|t|$ & hight-$|t|$    \\ \hline \hline

$M_{a_0}  (I)$   & 0.987 $\pm 0.002$   & 0.980 $\pm 0.005$    & 1.005 $\pm 0.01$     \\  \\
$M_{a_0}  (II)$               & 0.987 $\pm 0.002$   & 0.981 $\pm 0.007$    & 0.990 $\pm 0.006$   \\
\hline

$\Gamma_{a_0}(I)$  & 0.094 $\pm 0.011$          & 0.246  $\pm 0.031$         &0.37 $\pm 0.050$  \\ \\
$\Gamma_{a_0}(II)$   &  0.094 $\pm 0.010$     &  0.21 $\pm 0.04$   &  0.31 $\pm 0.04$   \\ \hline

$M_{a_2}(I)$   & 1.318 $\pm 0.003$   & 1.329 $\pm 0.002$    & 1.332 $\pm 0.003$     \\ \\
$M_{a_2}(II)$  & 1.318 $\pm 0.003$   & 1.329 $\pm 0.002$    & 1.335 $\pm 0.003$  \\
\hline

$\Gamma_{a_2}(I)$  &0.141 $\pm 0.0093$            & 0.155  $\pm 0.0067$         & 0.172 $\pm 0.0092$
\\ \\
$\Gamma_{a_2}(II)$    & 0.141 $\pm 0.009$  & 0.147 $\pm 0.007$  &  0.174 $\pm 0.009$  \\ \hline

$M_{X}(I)$ & 1.386  $\pm 0.032 $ & 9.99  $\pm 8.77$  & 3.47 $\pm 0.70$ &  \\ \\
$M_{X}(II)$               &   1.37      & 1.37      &  1.37 \\ \hline

$\Gamma_{X}(I)$& 0.363 $\pm 0.081$ &0.11 $\pm 0.31$ & 0.10 $\pm 0.27$  \\ \\
$\Gamma_{X}(II)$               &    0.37      &  0.37     & 0.37  \\ \hline

$\chi^2_{\nu} (I)$ &1.55 & 1.40 &  1.43  \\ \hline \hline

$\chi^2_{\nu} (II)$  &1.54 & 1.70 &  2.00  \\ \hline \hline

 \end{tabular}
 \caption{The mass and width for the $a_0(980)$, $a_2(1320)$ and a $P$-wave resonance
 obtained from a simultaneous fit to the
distributions in each of 12 moments.  Details are given in the text. For fit (I) the
$P$-wave mass and width are allowed to float.  For fit (II) the $P$-wave parameters are
fixed at the values reported in reference \cite{SUC1999}.  The $\chi^2$ per degree of
freedom, $\chi^2_{\nu}$ is also given for the two fits.}
\label{fits_mom}
\end{table}

\begin{figure}
\centerline{\epsfig{file=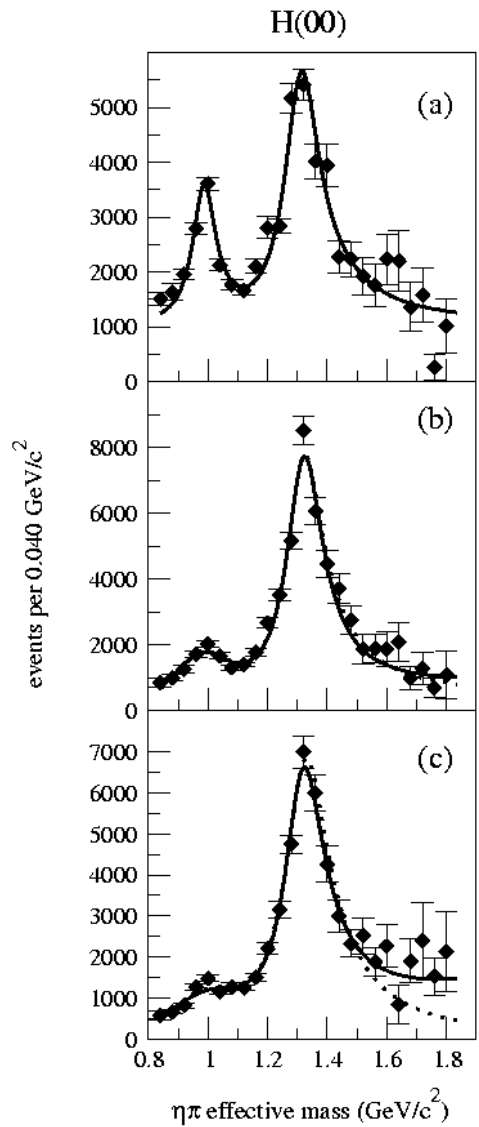,height=5.92in,width=2.5in}}
\caption{The observed $H(00)$ moment as a function of 
$\eta \pi^0$ effective mass for  different ranges of
momentum-transfer-squared, $|t_{\pi^{-}\to\eta\pi}|$:  (a) low $|t|$,
(b) medium $|t|$, and (c) high $|t|$.  The curves are results
of fits described in the text.
$H(00)=\left|{S_0}\right|^2+\left|{P_0}\right|^2+\left|{P_-}\right|^2+\left|{D_0}\right|^2+\left
|{D_-}\right|^2+\left|{P_+}\right|^2+\left|{D_+}\right|^2$ }
\label{H00}
\end{figure}

\begin{figure}
\centerline{\epsfig{file=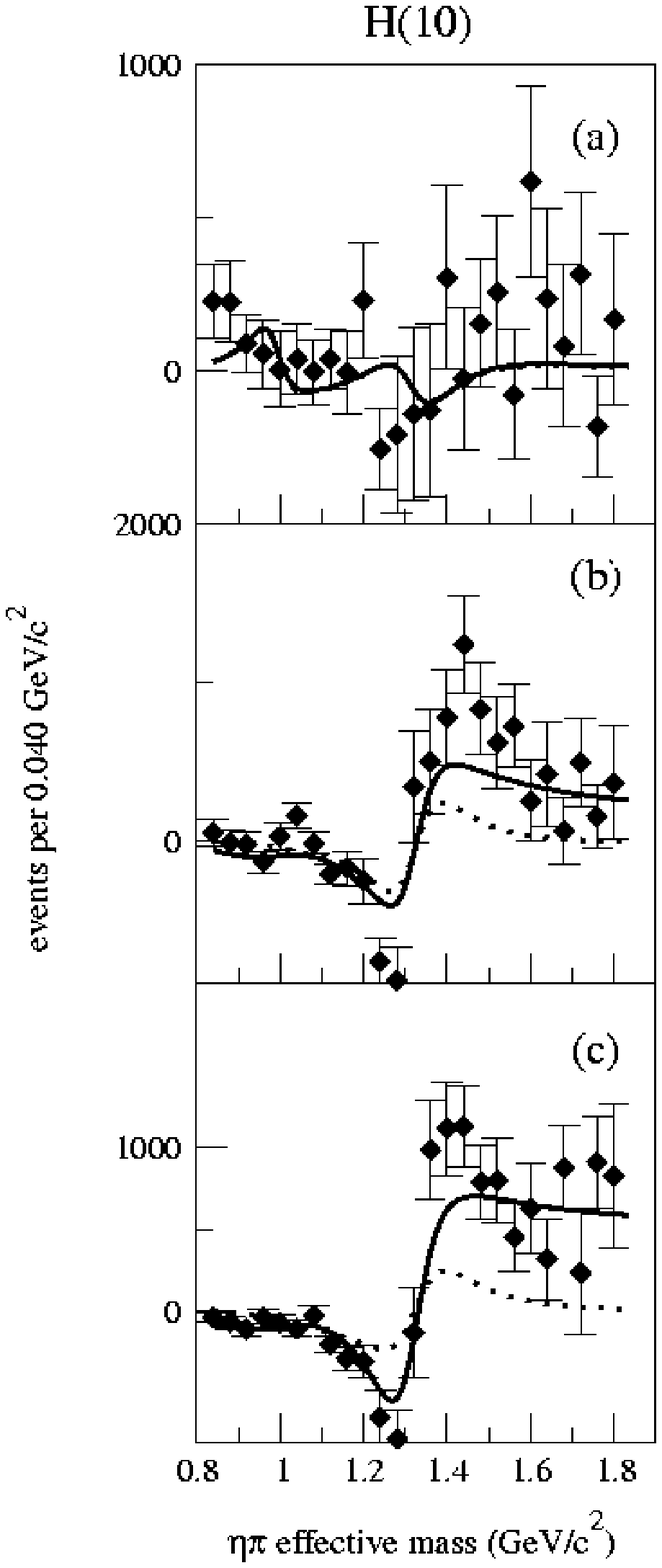,height=5.92in,width=2.5in}}
\caption{The observed $H(10)$ moment as a function of 
$\eta \pi^0$ effective mass for  different ranges of
momentum-transfer-squared, $|t_{\pi^{-}\to\eta\pi}|$:  (a) low $|t|$,
(b) medium $|t|$, and (c) high $|t|$.  The curves are results
of fits described in the text.
$H(10)={2 \over \sqrt{3}}Re\{S_0P_0^*\}+{4 \over \sqrt{15}}Re\{P_0D_0^*\}+{2 \over \sqrt{5}}Re\{P_-D_-^*\}+{2 \over
\sqrt{5}}Re\{P_+D_+^*\}$ }
\label{H10}
\end{figure}

\begin{figure}
\centerline{\epsfig{file=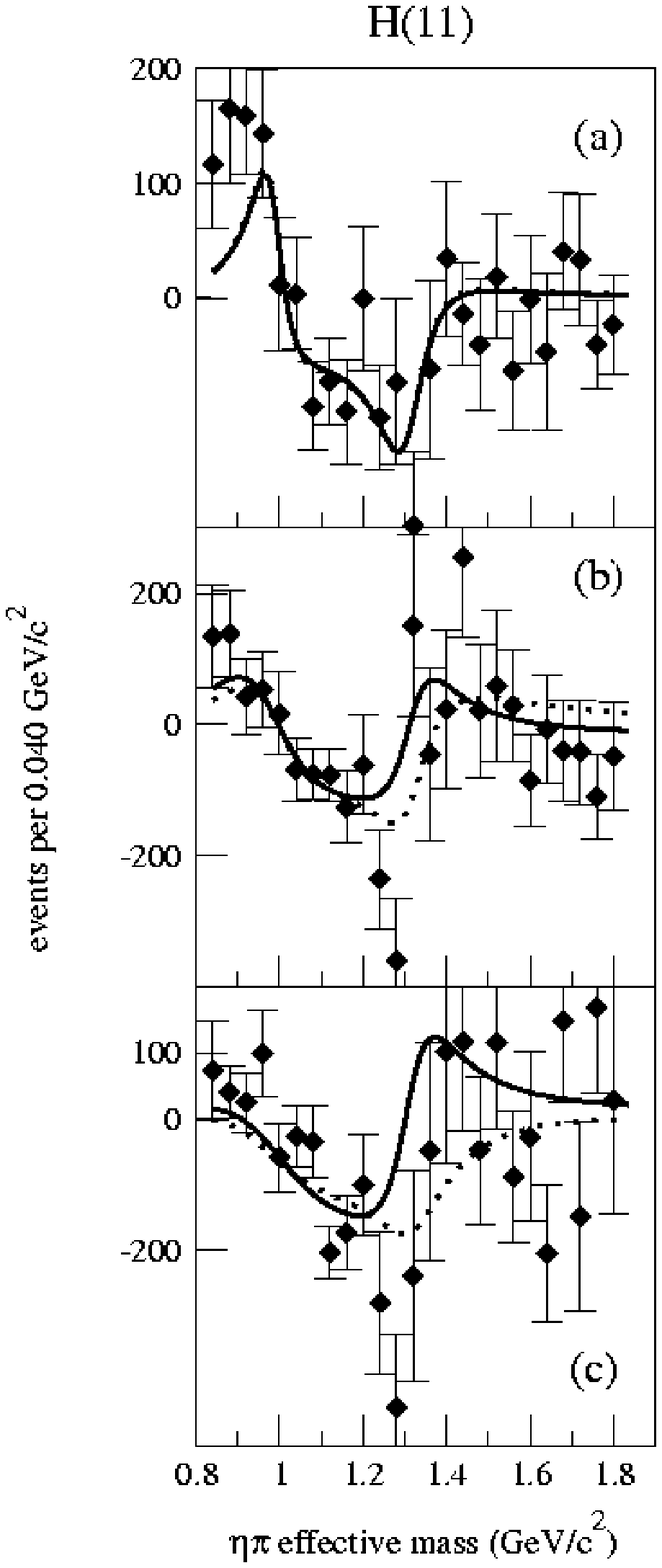,height=5.92in,width=2.5in}}
\caption{The observed $H(11)$ moment as a function of 
$\eta \pi^0$ effective mass for  different ranges of
momentum-transfer-squared, $|t_{\pi^{-}\to\eta\pi}|$:  (a) low $|t|$,
(b) medium $|t|$, and (c) high $|t|$.  The curves are results
of fits described in the text.
$H(11)={2 \over \sqrt{6}}Re\{S_0P_-^*\}+{2 \over \sqrt{10}}Re\{P_0D_-^*\}-{2 \over \sqrt{30}}Re\{P_-D_0^*\}$ }
\label{H11}
\end{figure}

\begin{figure}
\centerline{\epsfig{file=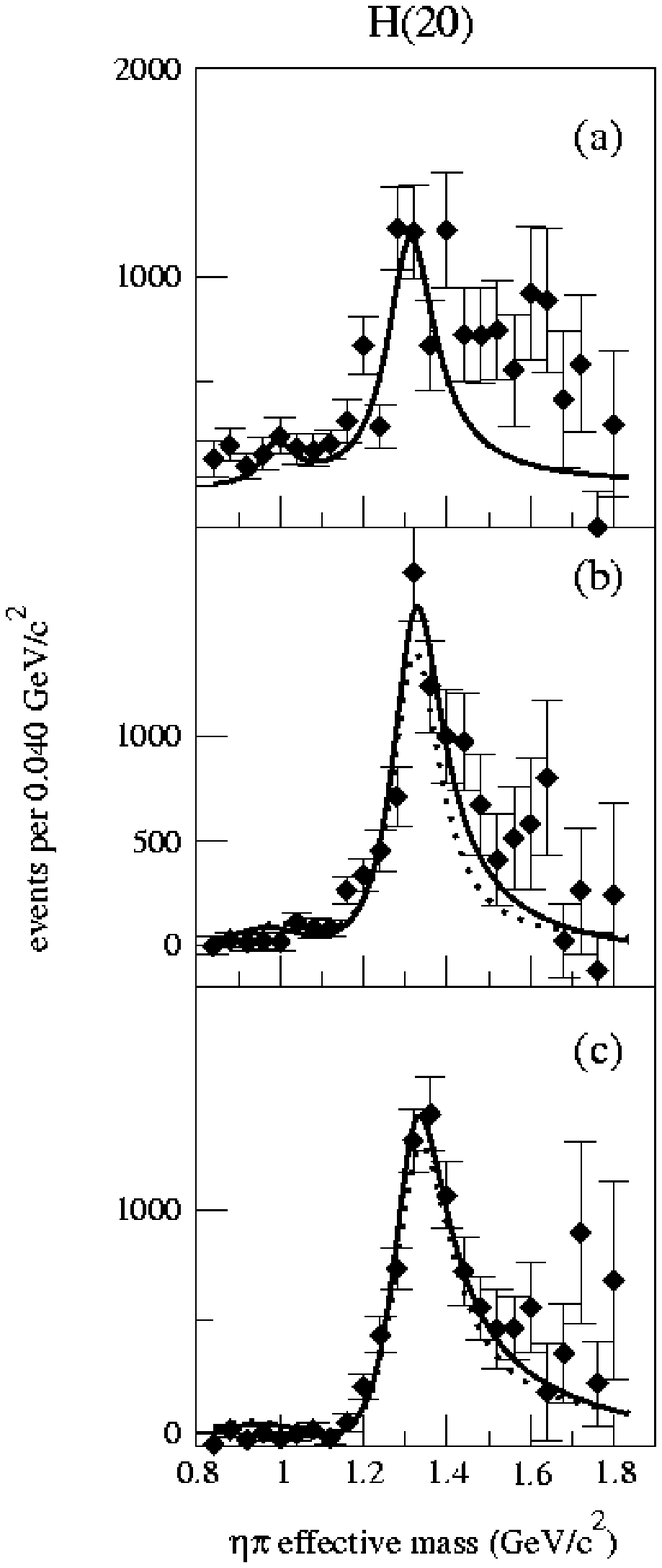,height=5.92in,width=2.5in}}
\caption{The observed $H(20)$ moment as a function of 
$\eta \pi^0$ effective mass for  different ranges of
momentum-transfer-squared, $|t_{\pi^{-}\to\eta\pi}|$:  (a) low $|t|$,
(b) medium $|t|$, and (c) high $|t|$.  The curves are results
of fits described in the text.
$H(20)={2 \over \sqrt{5}}Re\{S_0D_0^*\}+{2 \over 5}\left|{P_0}\right|^2-{1 \over 5}
\left|{P_-}\right|^2-{1 \over 5}\left|{P_+}\right|^2+{2 \over 7}\left|{D_0}\right|^2+{1 \over 7}
\left|{D_-}\right|^2+{1 \over 7}\left|{D_+}\right|^2$ }
\label{H20}
\end{figure}

\begin{figure}
\centerline{\epsfig{file=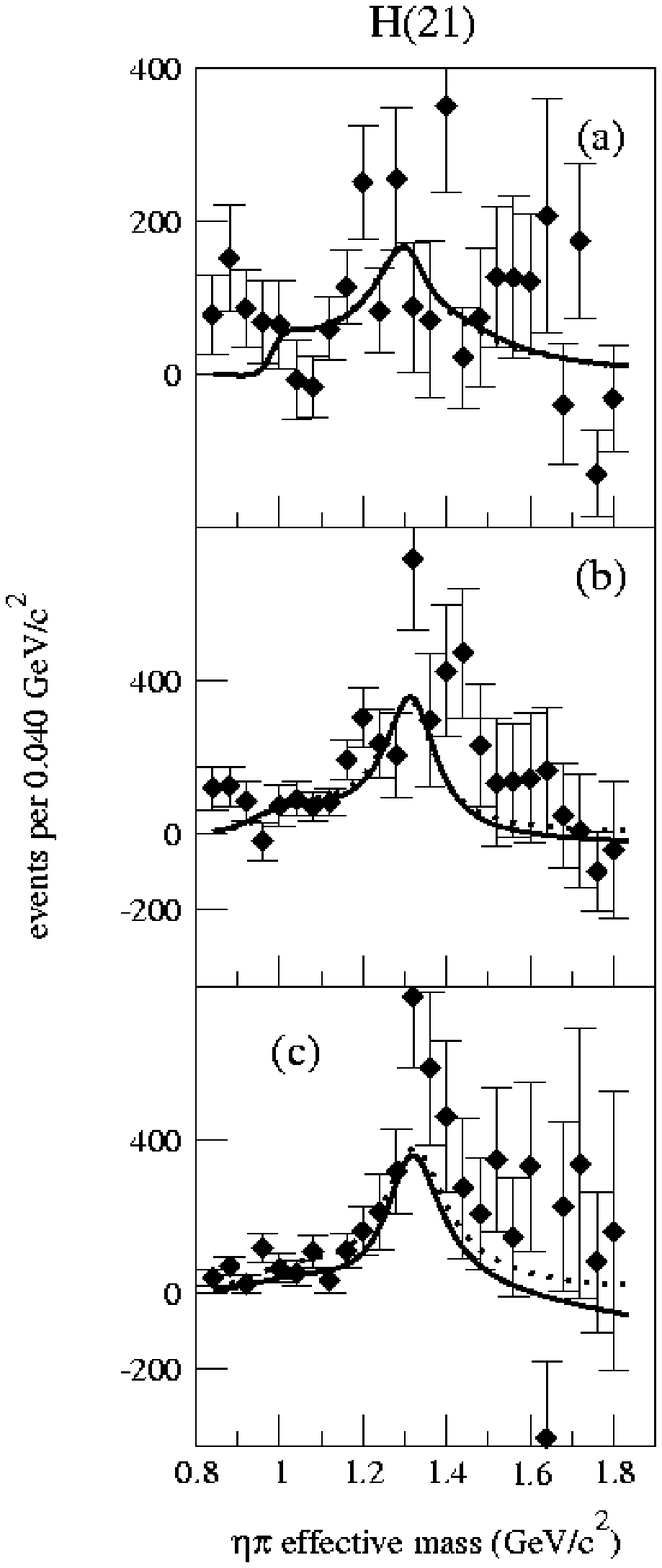,height=5.92in,width=2.5in}}
\caption{The observed $H(21)$ moment as a function of 
$\eta \pi^0$ effective mass for  different ranges of
momentum-transfer-squared, $|t_{\pi^{-}\to\eta\pi}|$:  (a) low $|t|$,
(b) medium $|t|$, and (c) high $|t|$.  The curves are results
of fits described in the text.
$H(21)={2 \over \sqrt{10}}Re\{S_0D_-^*\}+{2 \over 5}{\sqrt{3 \over 2}}Re\{P_0P_-^*\}+{2 \over 7\sqrt{2}}Re\{D_0D_-^*\}$ }
\label{H21}
\end{figure}

\begin{figure}
\centerline{\epsfig{file=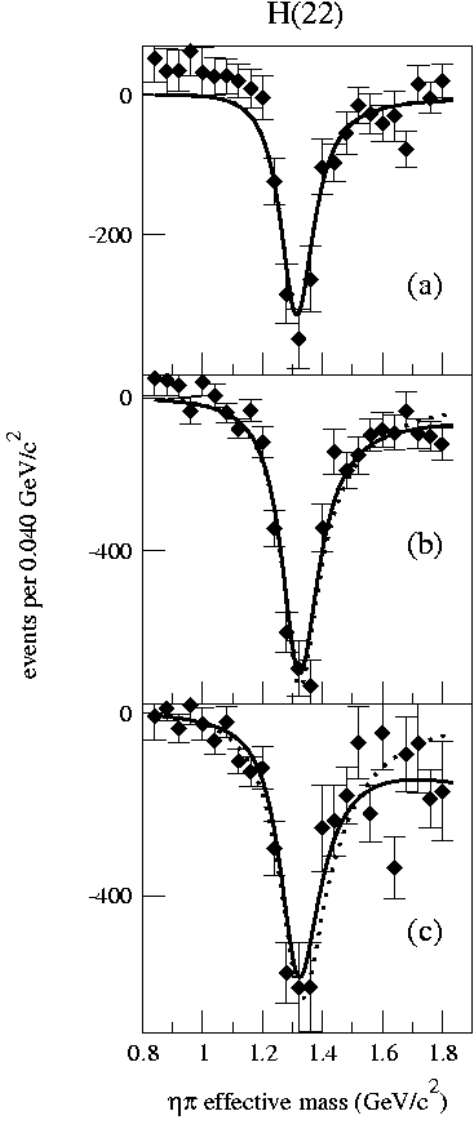,height=5.92in,width=2.5in}}
\caption{The observed $H(22)$ moment as a function of 
$\eta \pi^0$ effective mass for  different ranges of
momentum-transfer-squared, $|t_{\pi^{-}\to\eta\pi}|$:  (a) low $|t|$,
(b) medium $|t|$, and (c) high $|t|$.  The curves are results
of fits described in the text.
$H(22)={1 \over 5}{\sqrt{3\over2}}\left|{P_-}\right|^2-{1 \over 5}{\sqrt{3\over2}}\left|{P_+}
\right|^2+{1 \over 7}{\sqrt{3\over2}}\left|{D_-}\right|^2-{1 \over 7}{\sqrt{3\over2}}\left|{D_+}\right|^2$ }
\label{H22}
\end{figure}

\begin{figure}
\centerline{\epsfig{file=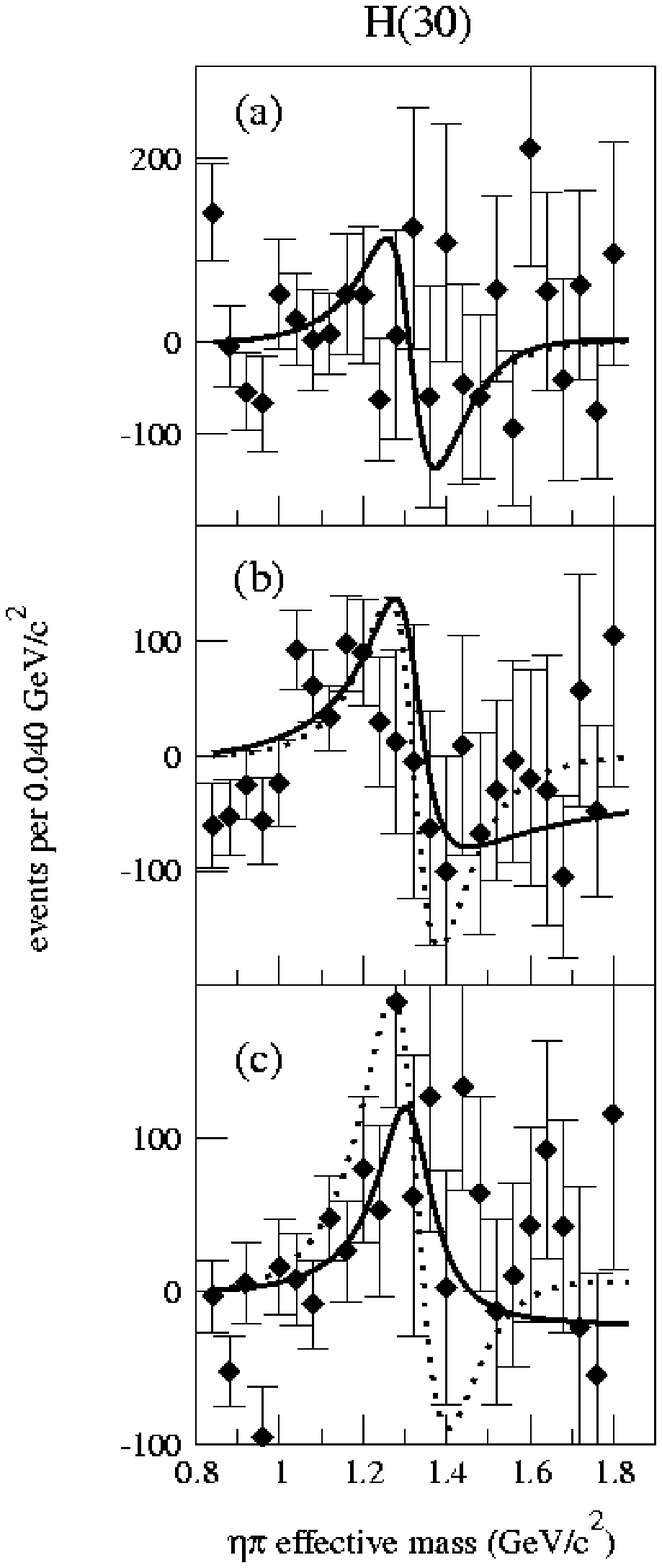,height=5.92in,width=2.5in}}
\caption{The observed $H(30)$ moment as a function of 
$\eta \pi^0$ effective mass for  different ranges of
momentum-transfer-squared, $|t_{\pi^{-}\to\eta\pi}|$:  (a) low $|t|$,
(b) medium $|t|$, and (c) high $|t|$.  The curves are results
of fits described in the text.
$H(30)={6\over7}{\sqrt{3\over5}}Re\{P_0D_0^*\}-{6 \over 7\sqrt{5}}Re\{P_-D_-^*\}-{6 \over 7\sqrt{5}}Re\{P_+D_+^*\}$ }
\label{H30}
\end{figure}

\begin{figure}
\centerline{\epsfig{file=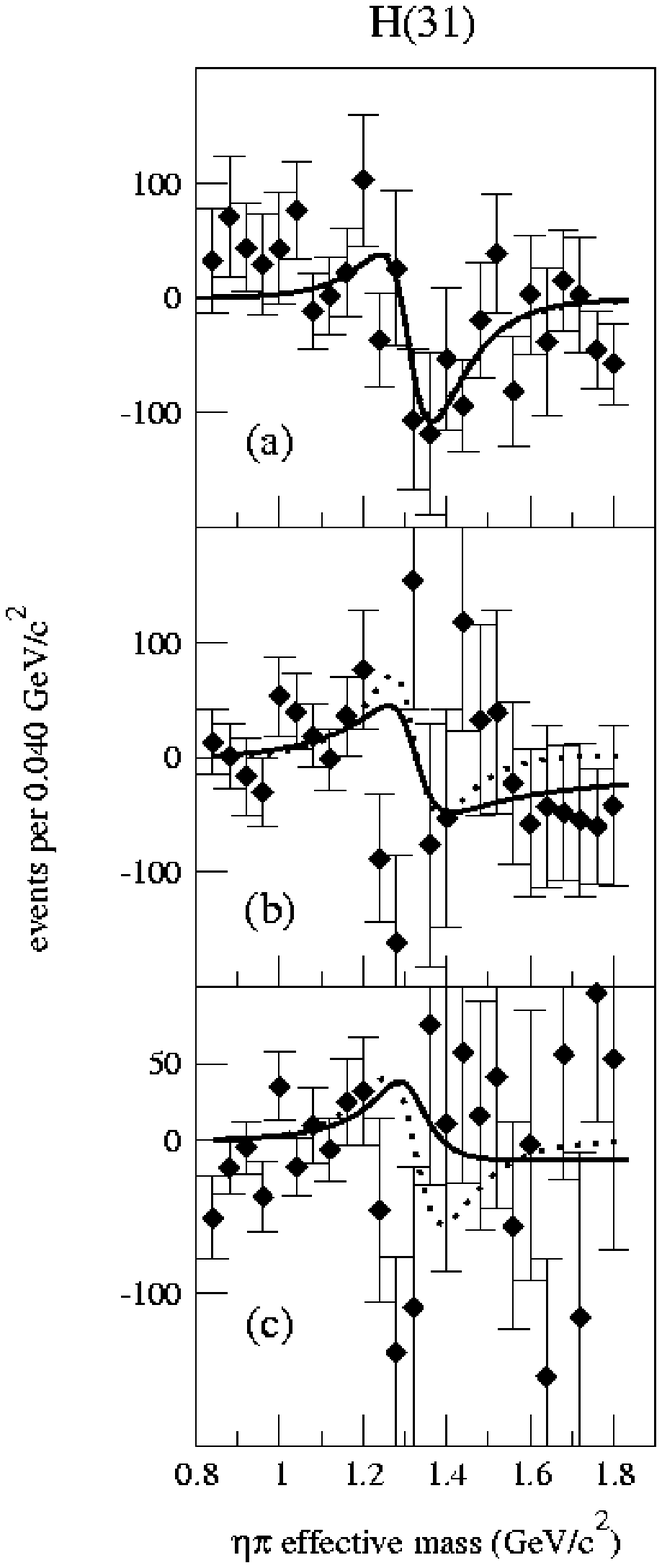,height=5.92in,width=2.5in}}
\caption{The observed $H(31)$ moment as a function of 
$\eta \pi^0$ effective mass for  different ranges of
momentum-transfer-squared, $|t_{\pi^{-}\to\eta\pi}|$:  (a) low $|t|$,
(b) medium $|t|$, and (c) high $|t|$.  The curves are results
of fits described in the text.
$H(31)={{4\over7}{\sqrt{3\over5}}}Re\{P_0D_-^*\}+{{6\over7\sqrt{5}}}Re\{P_-D_0^*\}$ }
\label{H31}
\end{figure}

\begin{figure}
\centerline{\epsfig{file=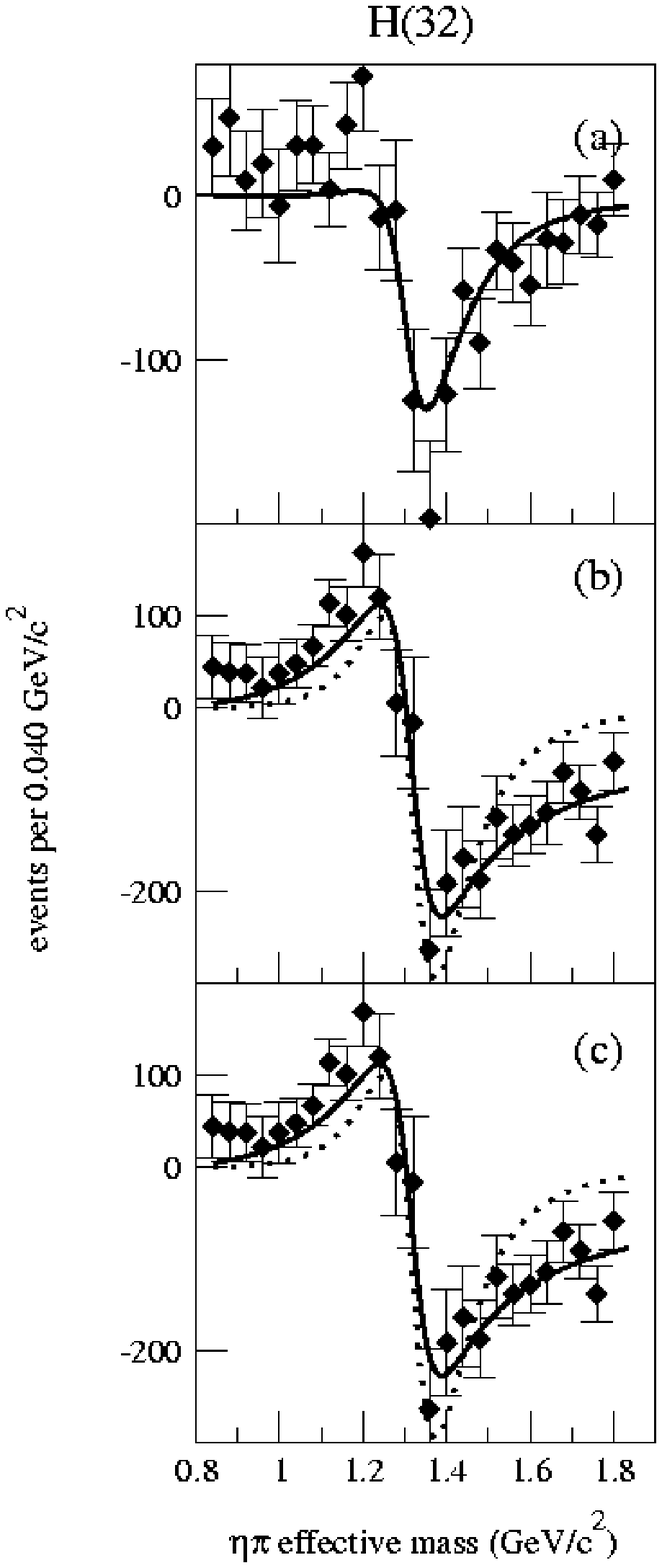,height=5.92in,width=2.5in}}
\caption{The observed $H(32)$ moment as a function of 
$\eta \pi^0$ effective mass for  different ranges of
momentum-transfer-squared, $|t_{\pi^{-}\to\eta\pi}|$:  (a) low $|t|$,
(b) medium $|t|$, and (c) high $|t|$.  The curves are results
of fits described in the text.
$H(32)={2\over7}{\sqrt{3\over2}}Re\{P_-D_-^*\}-{2\over7}{\sqrt{3\over2}}Re\{P_+D_+^*\}$ }
\label{H32}
\end{figure}

\begin{figure}
\centerline{\epsfig{file=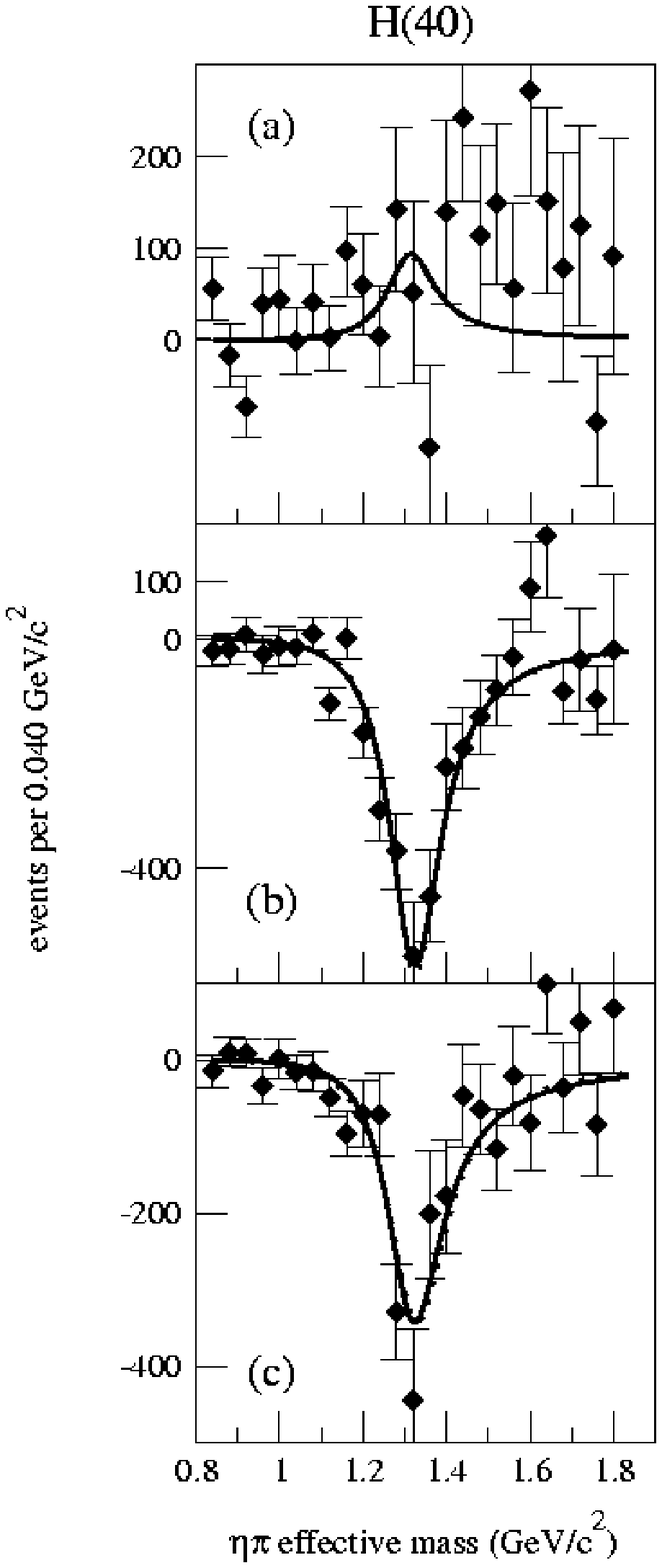,height=5.92in,width=2.5in}}
\caption{The observed $H(40)$ moment as a function of 
$\eta \pi^0$ effective mass for  different ranges of
momentum-transfer-squared, $|t_{\pi^{-}\to\eta\pi}|$:  (a) low $|t|$,
(b) medium $|t|$, and (c) high $|t|$.  The curves are results
of fits described in the text.
$H(40)={2\over7}\left|{D_0}\right|^2-{4\over21}\left|{D_-}\right|^2-{4\over21}\left|{D_+}\right|^2$ }
\label{H40}
\end{figure}

\begin{figure}
\centerline{\epsfig{file=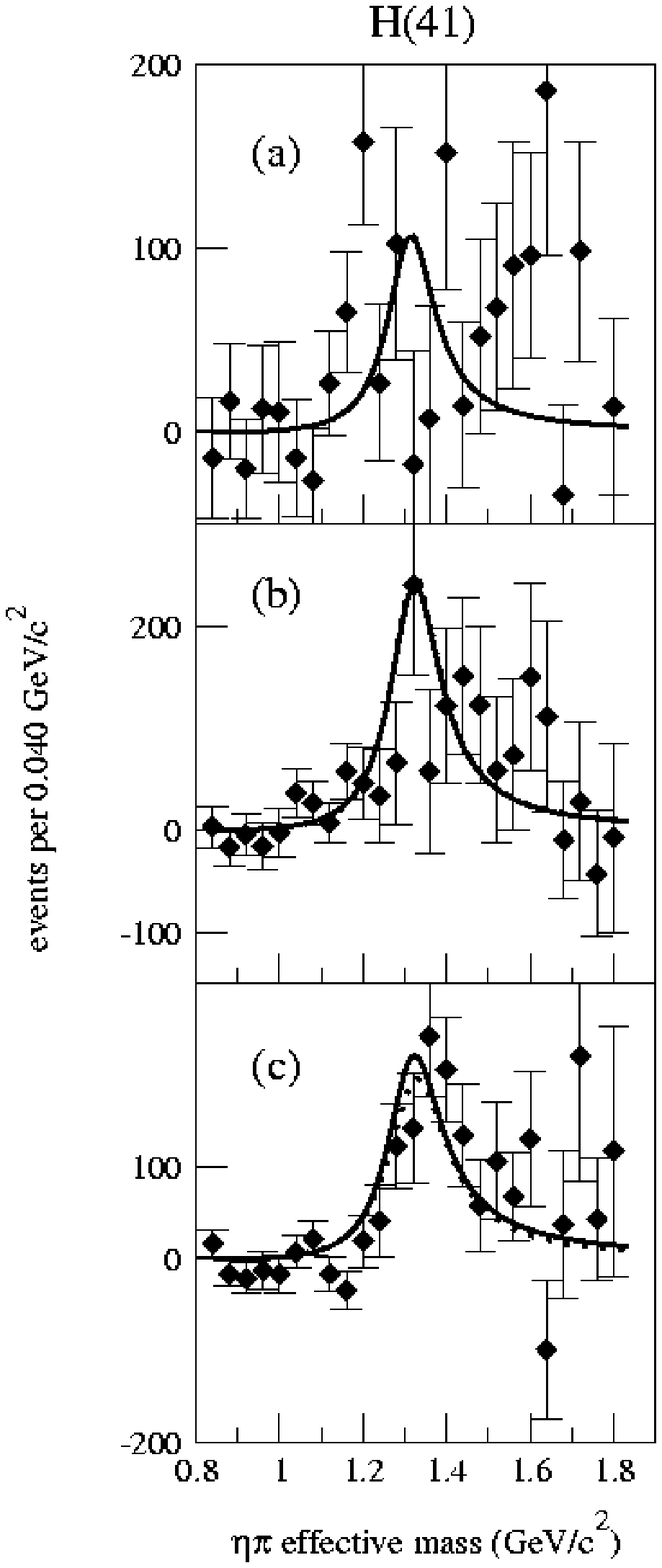,height=5.92in,width=2.5in}}
\caption{The observed $H(41)$ moment as a function of 
$\eta \pi^0$ effective mass for  different ranges of
momentum-transfer-squared, $|t_{\pi^{-}\to\eta\pi}|$:  (a) low $|t|$,
(b) medium $|t|$, and (c) high $|t|$.  The curves are results
of fits described in the text.
$H(41)={2\over7}{\sqrt{5\over3}}Re\{D_0D_-^*\}$ }
\label{H41}
\end{figure}

\begin{figure}
\centerline{\epsfig{file=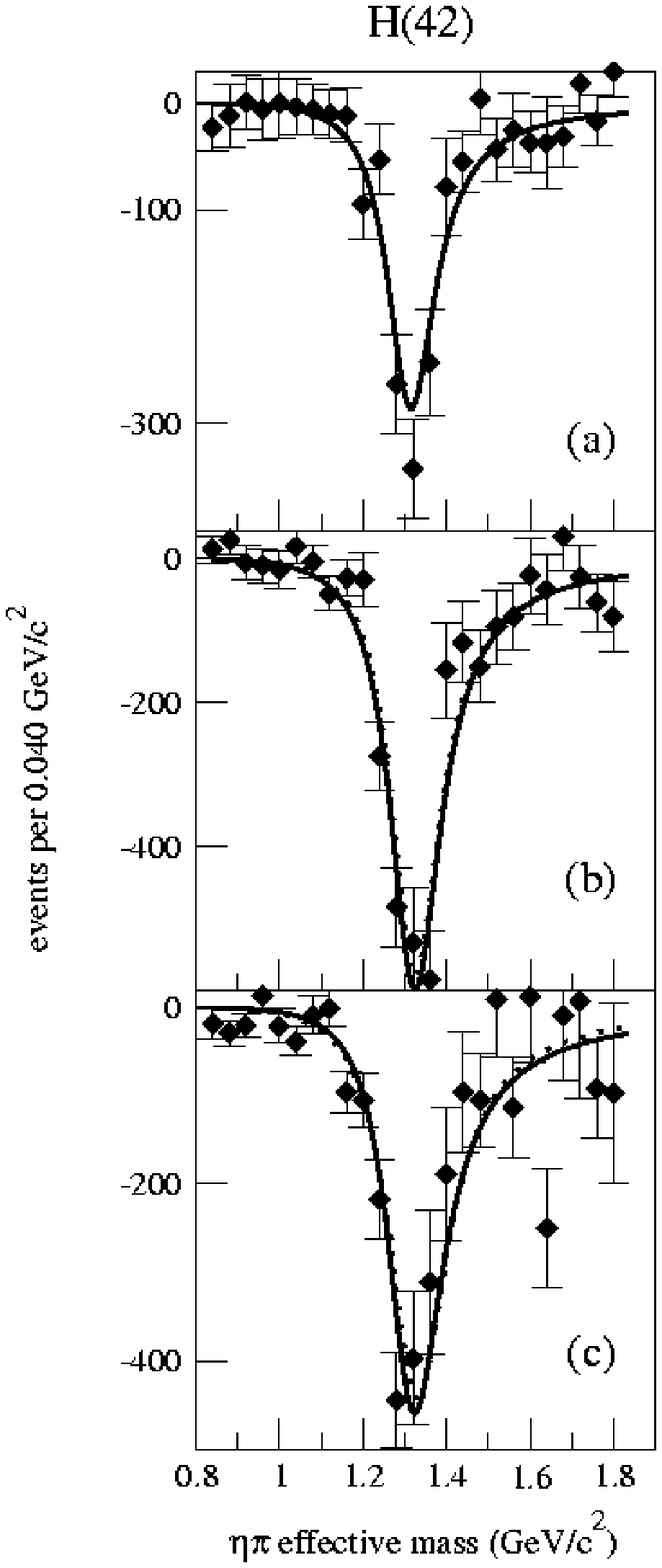,height=5.92in,width=2.5in}}
\caption{The observed $H(42)$ moment as a function of 
$\eta \pi^0$ effective mass for  different ranges of
momentum-transfer-squared, $|t_{\pi^{-}\to\eta\pi}|$:  (a) low $|t|$,
(b) medium $|t|$, and (c) high $|t|$.  The curves are results
of fits described in the text.
$H(42)={\sqrt{10}\over21}\left|{D_-}\right|^2-{\sqrt{10}\over21}\left|{D_+}\right|^2$ }
\label{H42}
\end{figure}

The outcome of the fitting is shown for 
all moments
for the three ranges of 
$t$ described above in figures \ref{H00} through \ref{H42}. 
The moments shown are corrected for acceptance using the procedure described in
reference \cite{Chu97}.
 The solid curves are
the result of fits in which all of the 21 parameters described above were allowed to float.  The
dashed curves are the results of fits where the $P$-wave Breit-Wigner 
 mass was fixed at 1.37~GeV$/c^2$\ and the the width was fixed
at 0.37~GeV$/c^2$.  These are the parameters reported for the $P$-wave exotic in
the $\eta \pi^-$ channel in reference~\cite{SUC1999}.

In the low-$t$ region, the $S_0$ wave is prominent at
  $m_{\eta\pi^0} \sim$ 1 GeV$/c^2$\ 
 and agrees well with the expected contribution to the $\eta\pi^0$
 spectrum from the decay of the $a_0(980)$ meson. The $S_0$-wave is 
produced 
 without flipping helicities at the meson vertex and therefore it is 
 dominant at low-$t$ and decreases with increasing $|t|$. 
 The structure at  $m_{\eta\pi^0} \sim$ 1~GeV$/c^2$\  in the 
 $H(10)$ and $H(11)$ moments (figures \ref{H10} and 
\ref{H11} respectively) reflects the interference between the $S$
  and $P$ waves and it is noticeable only at low-$t$ where the $S$-wave
  production is the strongest.  
  The  $D_+$ wave dominates the spectrum at
  $m_{\eta\pi^0} \sim$ 1.3~GeV$/c^2$\ and since it is produced  with a 
meson
 helicity flip it is suppressed at low-$t$ and it is the largest in the
   medium-$t$ region.  In all three $t$-regions, however it 
 is  well described by a Breit-Wigner shape corresponding to
 the $a_2(1320)$ meson. 
   As the $D_+$ production increases (to medium and large $t$-regions), 
 the structure in the moment $H(10)$ (figure \ref{H10})
  moves to the $a_2$ region and a similar structure in the moment 
  $H(30)$ (figure \ref{H30}) appears. Similarly, the interference of the negative
  reflectivity $P$-waves with the negative reflectivity $D$-waves 
 is reflected in the moments $H(11)$ and $H(31)$ (figures \ref{H11} and \ref{H31}
respectively)
at $m_{\eta\pi^0} \sim$ 1.3~GeV$/c^2$. 

The fit corresponding to the fixed $P$-wave parameters (dashed curve)
is significantly worse in $H(10)$ (figure \ref{H10}) -- a moment linear
in the $P$-wave and somewhat worse or no better in the
  other moments that are linear in the 
$P$-wave --  $H(11)$ (figure \ref{H11}), 
$H(30)$ (figure \ref{H30}) and
 $H(31)$ (figure \ref{H31}).
In addition to the $H(00)$ moment distribution, the other moments -- 
 $H(20)$, $H(22)$, $H(40)$ and $H(42)$ -- are primarily sensitive to the 
 presence of the  $a_2$ meson, and therefore are not very
 sensitive to treatment of $P$-waves in the fit. 
  The fit with the $P$-wave parameters fixed gives a
  significantly worse $\chi^2$ per degree of freedom ($\chi^2_\nu$ from 1.70 to 2.00)
  as compared to 
 the case when the $P$-wave parameters are allowed to float ($\chi^2_\nu$ from 1.37 to 1.43). 
 In the later case, however, it was found  that no single set of 
consistent parameters 
 could be found for the
 $P$ waves across the  three $t$-ranges. In the low-$t$ range the $P$-wave 
 with mass of 1.39 GeV$/c^2$\ and width of 0.36 GeV$/c^2$\ was
 found, which is the only solution  consistent with what was reported for
 the $\eta\pi^-$ channel \cite{SUC1999}. 
  However, even in this $t$-range other solutions 
 could also be found with masses ranging  from 1 to 5 GeV$/c^2$.
 For example, a solution for a $P$-wave with a mass of
0.91 GeV$/c^2$\ and a width of 1.05 GeV$/c^2$\ had a $\chi^2_{\nu} =
1.57$.  This is because the interference with the $D$-waves is
weak at low-$t$ and the $S$-wave is too narrow to constrain the $P$ away from the
 $m_{\eta\pi^0} \sim$ 1.3 GeV$/c^2$\  region. 
The medium-$t$ to large-$t$ range should be the best for constraining
the $P$  waves. Here the $D$ waves are  dominant and some $S$-wave is 
 still present. In this $t$-range, however,
 acceptable solutions (with comparable $\chi^2_\nu$) different from the one
  listed in table \ref{fits_mom} could not be found.

The non-vanishing  moments $H(10)$ (figure \ref{H10}), 
 $H(11)$ (figure \ref{H11}), 
$H(30)$ (figure \ref{H30}), 
$H(31)$ (figure \ref{H31}) and $H(32)$ (figure \ref{H32}) clearly indicate a presence of
 a  $P$-wave in the $\eta\pi^0$ system.  The large values of mass and width
 found indicate  that the Breit-Wigner   parameterization for 
 these waves is most likely inadequate.

\section{Mass and $t$ Dependence of Partial Waves}

\subsection{PWA solutions and fits to the moments}

\begin{figure}
\centerline{\epsfig{file=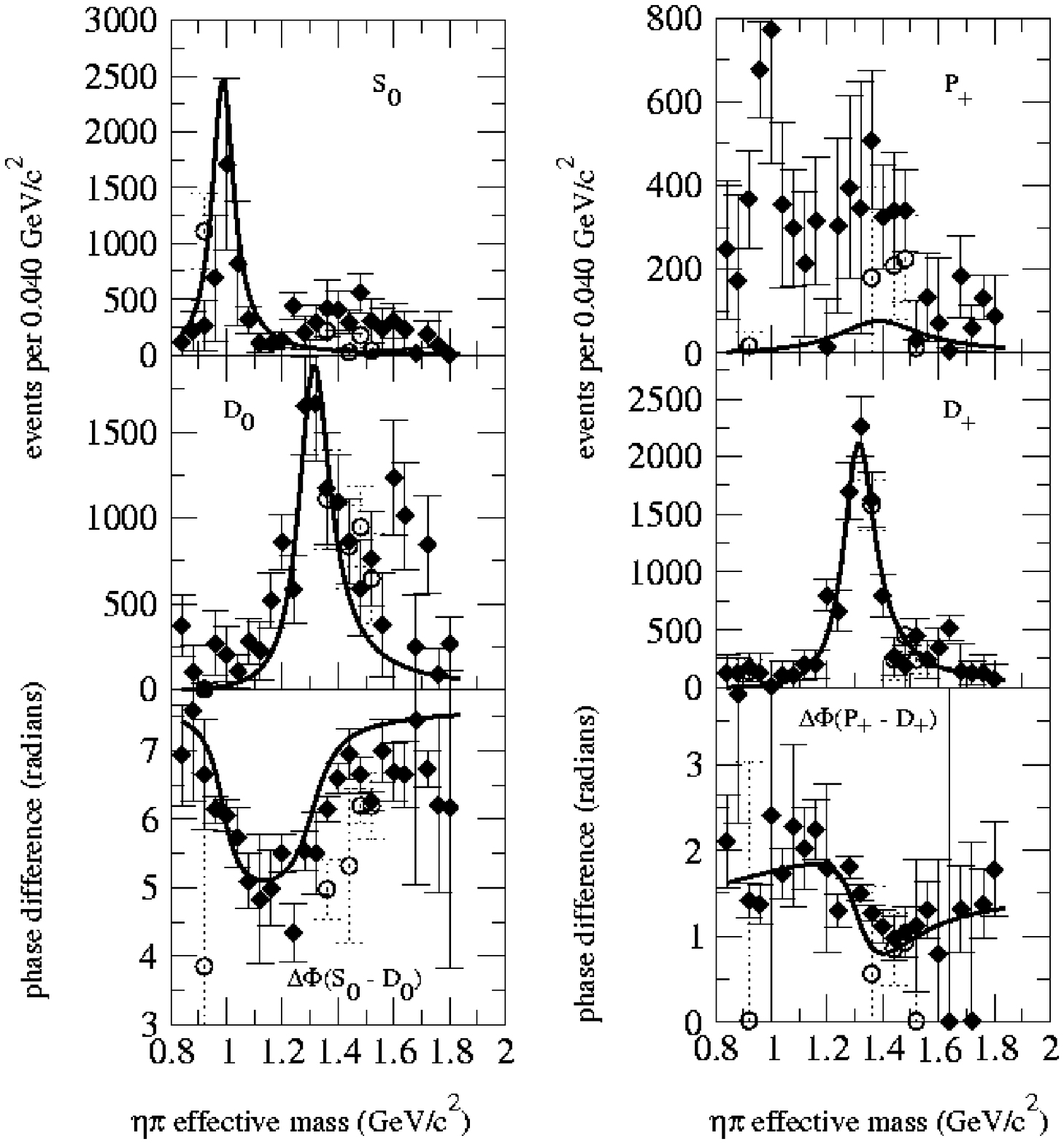,width=\columnwidth}}
\caption{PWA solutions for the:
(left) -- (top) $S_0$ (middle) $D_0$ waves
and (bottom) $\Delta \Phi(S_0 - D_0)$ phase difference and for the 
(right) -- (top) $P_+$ (middle) $D_+$ waves
and (bottom) $\Delta \Phi(P_+ - D_+)$ phase difference for the low-$|t|$ region.
The fits are result of the fit to the moments. The filled diamonds correspond
to the selected solution.
The open circles are additional
solutions allowed by the criteria described in the text.}
\label{mft_stk_t1}
\end{figure}

\begin{figure}
\centerline{\epsfig{file=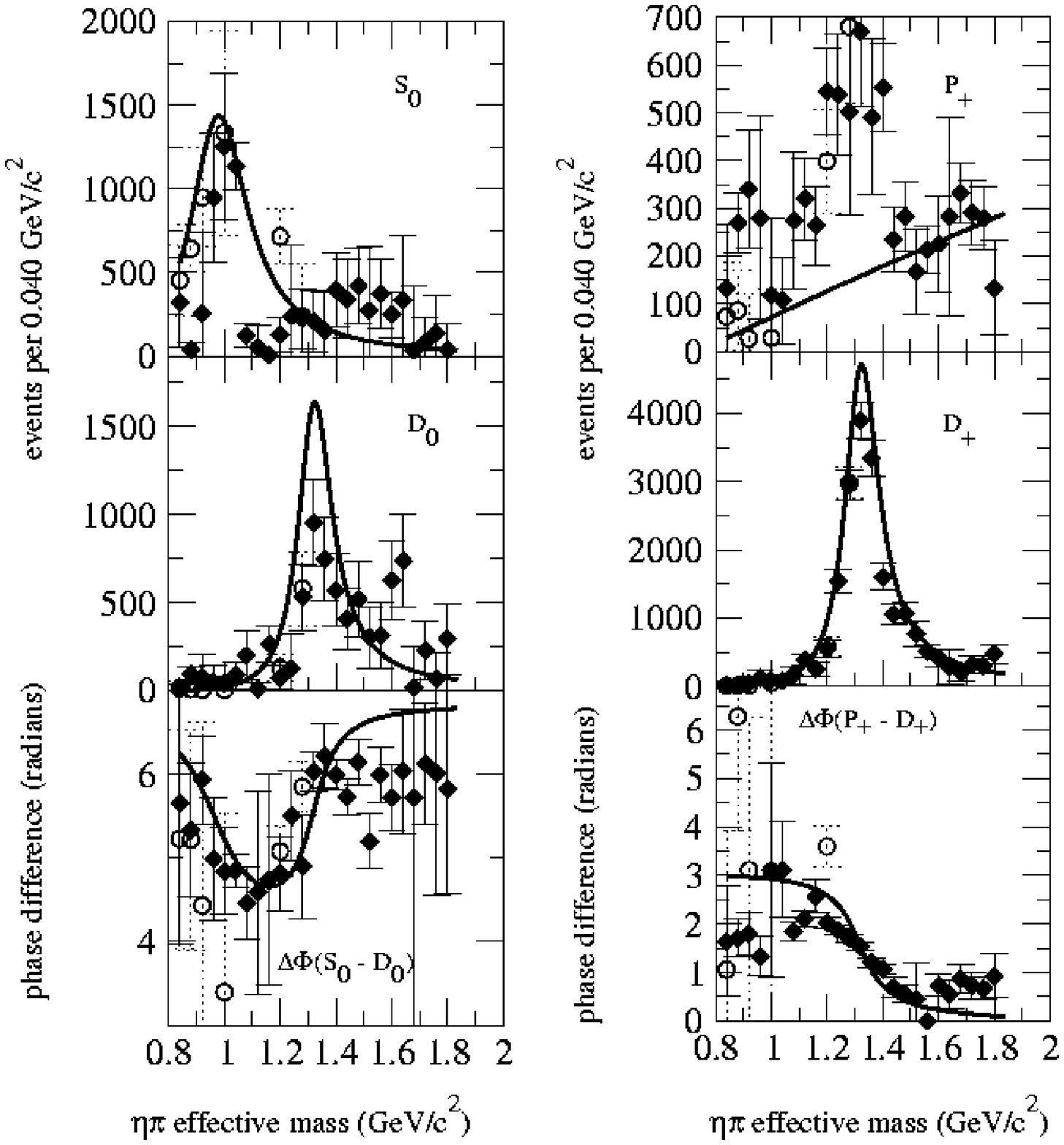,width=\columnwidth}}
\caption{PWA solutions for the:
(left) -- (top) $S_0$ (middle) $D_0$ waves
and (bottom) $\Delta \Phi(S_0 - D_0)$ phase difference and for the 
(right) -- (top) $P_+$ (middle) $D_+$ waves
and (bottom) $\Delta \Phi(P_+ - D_+)$ phase difference for the medium-$|t|$ region.
The fits are result of the fit to the moments. The filled diamonds correspond
to the selected solution.
The open circles are additional
solutions allowed by the criteria described in the text.}
\label{mft_stk_t2}
\end{figure}

\begin{figure}
\centerline{\epsfig{file=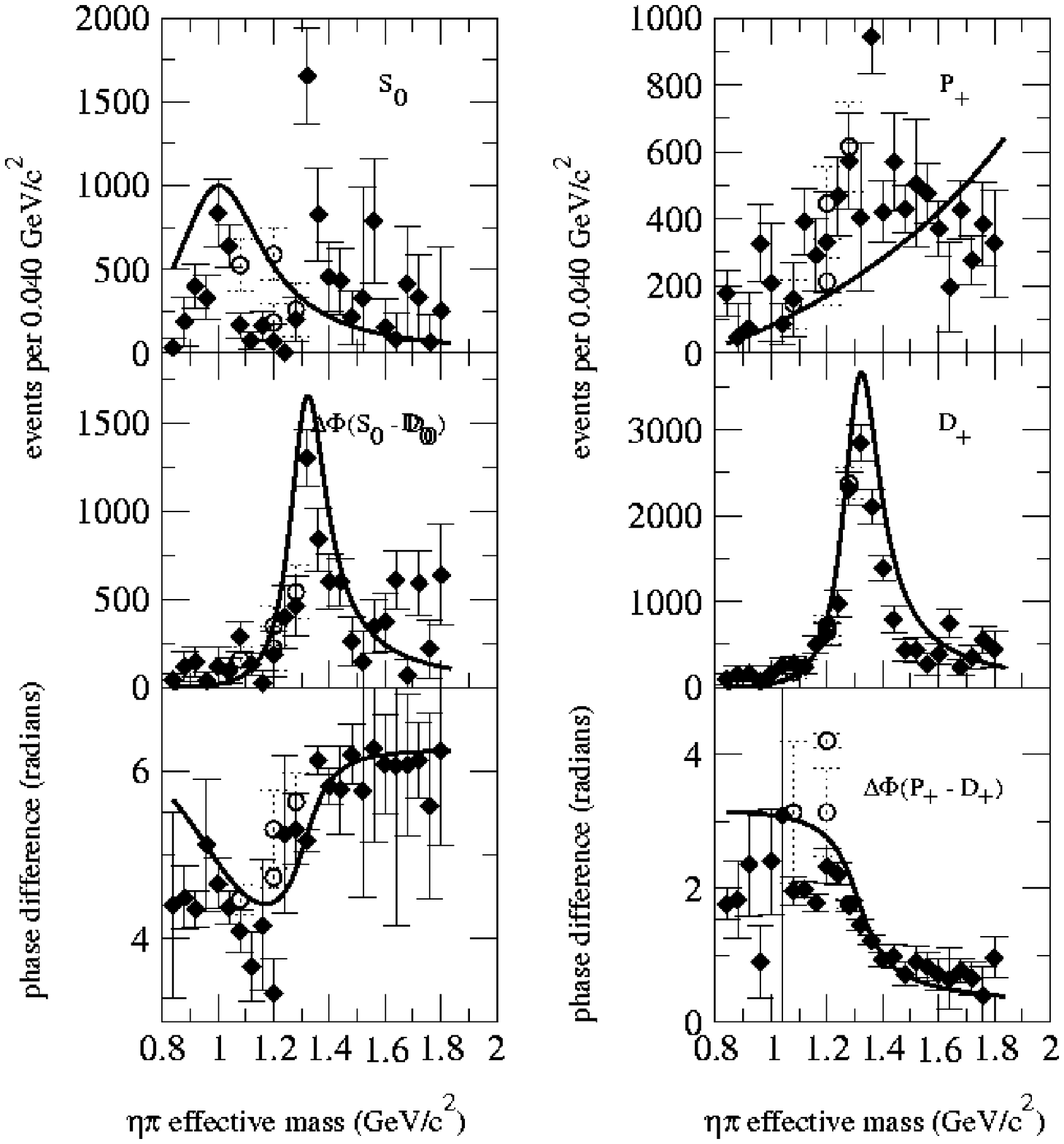,width=\columnwidth}}
\caption{PWA solutions for the:
(left) -- (top) $S_0$ (middle) $D_0$ waves
and (bottom) $\Delta \Phi(S_0 - D_0)$ phase difference and for the 
(right) -- (top) $P_+$ (middle) $D_+$ waves
and (bottom) $\Delta \Phi(P_+ - D_+)$ phase difference for the high-$|t|$ region.
The fits are result of the fit to the moments. The filled diamonds correspond
to the selected solution.
The open circles are additional
solutions allowed by the criteria described in the text.}
\label{mft_stk_t3}
\end{figure}

In figures~\ref{mft_stk_t1} through \ref{mft_stk_t3} various partial waves and
phase differences are shown as a function of $\eta \pi^0$ mass for the 
low-$|t|$, medium-$|t|$ and high-$|t|$ ranges.  In particular we show the
mass dependence of the $S_0$ and $D_0$ amplitudes and the $\Delta \Phi (S_0 - D_0)$
phase difference. 
 Also shown is the
mass dependence of the $P_+$ and $D_+$ amplitudes and the $\Delta \Phi (P_+ - D_+)$
phase difference. 
The filled data points indicate the selected (physical) solutions.  The additional points indicated by open circles
and dashed error bars are those ambiguous solutions that satisfy conditions described 
below. 

 The curves are the result of the fits to the moments 
 where the $P$-wave Breit-Wigner parameters are allowed to float.  
 We refer the reader to similar plots shown in figure~\ref{criteria}
for the  $S_0$ and $D_+$ amplitudes and the $\Delta \Phi (S_0 - D_0)$
phase difference along the the curves used to select the solutions.

The results of the fits to the moments (recalling that the moments are unambiguous)
  were used to study the possible inclusion of other
ambiguous solutions not selected as the physical solutions. 
 We compute a $\chi^2$ that measures the deviation of 
other ambiguous solutions from the partial wave solutions obtained from the moments
fits.  This requirement was imposed separately for the $S_0$ and $D_+$-waves 
and the $\Delta \Phi(S_0 - D_0)$ phase difference. 
In some cases  several
ambiguous solutions were found to have $\chi^2$ similar to the selected solutions
 and these are included in figures
figures~\ref{mft_stk_t1} through \ref{mft_stk_t3} as the open circles with dashed error bars.  
 The mass-dependent fits involving
only the $P_+$ and $D_+$ amplitudes and the $\Delta \Phi (P_+ - D_+)$
-- as described below --
phase difference were  insensitive to the inclusion of these solutions.

The fits shown in  figures~\ref{mft_stk_t1} through \ref{mft_stk_t3} describe
the $D_0$ and $D_+$ in all three $|t|$ regions.  The values obtained for the mass
and width of the $a_2(1320)$ as shown in table \ref{fits_mom} are consistent with
nominal values for this well understood state.  The $S_0$ is well described in the
low-$|t|$ region and the mass and width shown in table \ref{fits_mom} for the $a_0(980)$
agree with published values. As $|t|$ increases the production of the $a_0(980)$ decreases
and the description of the $S_0$ wave in figure~\ref{mft_stk_t3}, for the high-$|t|$
range is problematic. In this region the $S_0$ wave includes mostly structureless
background naturally subsumed into the $S_0$ wave.  Requiring that the
$S_0$ wave be described by a Breit-wigner wave yields a width that is wide and
driven by the phase motion of the $a_2(1320)$ via the $\Delta \Phi(S_0 - D_0)$ phase difference.
As noted above, the moment fits  did not constrain
 the parameters of a possible second scalar resonance, the $a_0(1450)$.
In the mass-dependent amplitude and phase difference fits, the inclusion
of a second scalar resonance
 In the low-$t$ region  yields $M= 1.34 \pm 0.06
 \mbox{ GeV}$
 and $\Gamma= 0.41 \pm 0.09  \mbox{ GeV}$ for the
 mass and width of the second $a_0$ resonance, while in the medium-$t$
region  we obtain $M =1.44 \pm 0.03 \mbox{ GeV}$ and $\Gamma = 0.25
\pm 0.06\mbox{ GeV}$, respectively. No stable fit was found for
 high-$t$ data.

The moments fits poorly describe the $P_+$ waves in all three $|t|$ regions
and the $P_+$ amplitude peaks in the vicinity of the dominant $D_+$ wave strongly
suggesting leakage.  We also note that the $P_+$ wave has an additional lower mass peak 
in the vicinity of the $a_0(980)$ in the low-$|t|$ region (see figures~\ref{mft_stk_t1} 
and  \ref{mft_stk_t2} where production of the $a_0(980)$ is strong -- again suggesting 
leakage.  Our Monte Carlo studies indicate that leakage of dominant waves can lead to false 
peaks in other amplitudes but it is unlikely to cause false phase motion.   Indeed this 
$P_+$ wave enhancement at about the mass of the $a_0(980)$  follows closely the behavior of 
the $S$ wave in its dependence on both mass and  momentum transfer squared.
Furthermore as shown in figure~\ref{mft_stk_t1} and especially in figure~\ref{mft_stk_t2}  
the additional solutions allowed by the selection criteria (those indicated by open circles)
do not alter the $S$ wave shape but reduce the amplitude of the $P_+$ wave in the region of 
$\sim 1$~GeV/$c^2$. Furthermore, the fits to the moments do not reveal any $P$ wave signal 
at $\sim 1$~GeV/$c^2$ and  due to angular momentum barrier effects it is expected that at low
breakup momentum the $S$ wave should dominate over the $P$ wave.

\subsection{Systematics and the fitting procedure}

\begin{figure}
\centerline{\epsfig{file=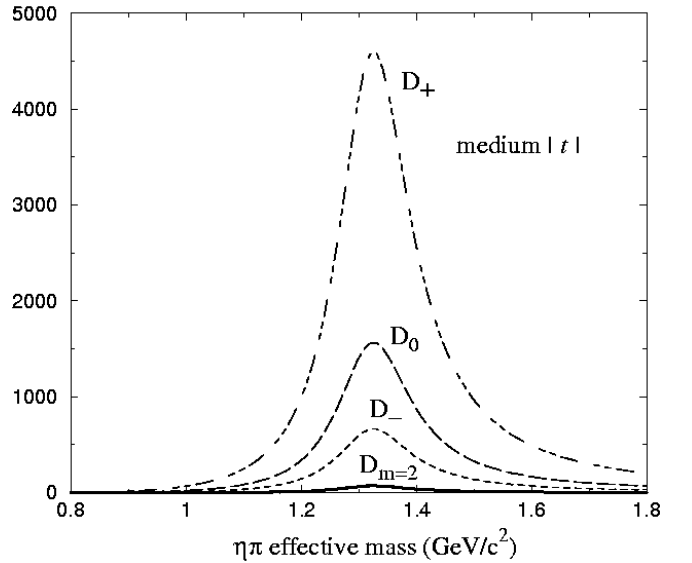,width=\columnwidth}}
\caption{Comparison of the intensity of the $D$ waves for the
medium-$|t|$ region including the $D$ waves
with $M$=2. There are two $M$=2 waves with opposite reflectivity. }
\label{dwaves}
\end{figure}

The assumption was made earlier that waves with $|M| > 1$ were not important.  In order
to test this assumption the additional $|M| = 2$ $D$ waves, one for each reflectivity,
were included in the fits for all the moments including
 the $H(43)$ and $H(44)$ moments.  The Breit-Wigner distributions
corresponding to the resulting fits are shown in figure~\ref{dwaves} for the 
$D_+$, $D_0$, $D_-$ waves and the two $D_{M=2}$ waves of opposite reflectivities (the
two $M=2$ curves are indistinguishable in the plot).  This supports our original assumption.

\begin{figure}
\centerline{\epsfig{file=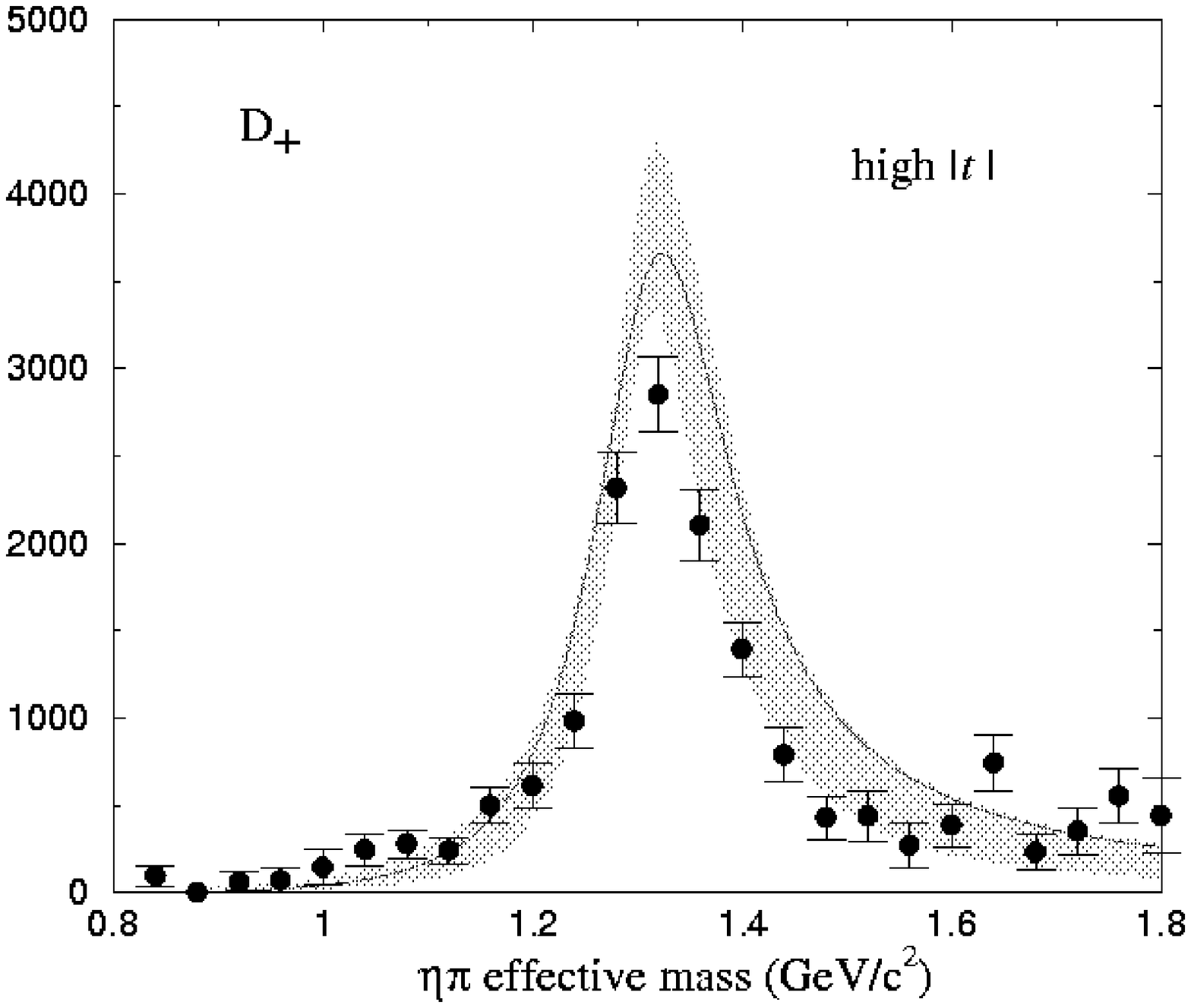,width=\columnwidth}}
\caption{The $D_+$ wave as a function of $\eta \pi^0$ effective mass for the
high-$|t|$ range.  The curve is the result of  an overall
fit to the moments, including a $D$-wave resonance,
as described in the text.  The shaded area is an estimate of the systematic
errors derived from a variation of
 all combinations of the Breit-Wigner parameters for the
 $S$, $P$ and $D$ amplitudes -- as described in the text. }
\label{systematics}
\end{figure}

The systematics of varying all possible fit assumptions was studied for the $D_+$-wave
for the high-$|t|$ region and the results are shown in the plot of figure~\ref{systematics}.
We chose to study this high-$|t|$ region since the description of the $D_+$ wave
amplitude is poorest for this region ((see figure~\ref{mft_stk_t3}).
  The curve is the result of a of the overall
fit to the moments.  The shaded region corresponds to the variations in the fit by letting
the Breit-Wigner parameters for all the  other waves vary in all possible combinations --
that is, being fixed at nominal values or allowed to float.  The $D_+$ wave is within the
systematic errors.

\subsection{Mass-dependent fits}

\begin{figure}
\centerline{\epsfig{file=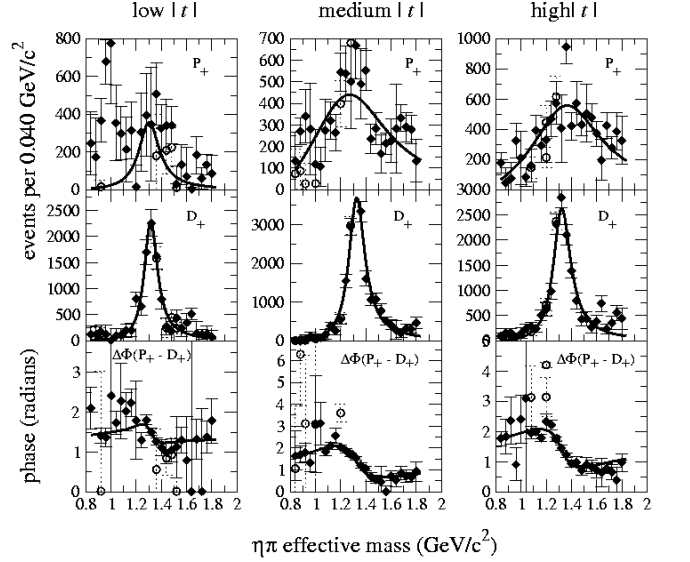,width=\columnwidth}}
\caption{PWA solution for the (top) $P_+$ (middle) $D_+$ waves
and (bottom) $\Delta \Phi(P_+ - D_+)$ phase difference for
low-$|t|$, medium-$|t|$ and high-$|t|$ as a function of 
$\eta \pi^0$ effective mass. The curves are the result of a mass-dependent
fit assuming Breit-Wigner resonance forms for the $P_+$ and $D_+$ waves. }
\label{massdep}
\end{figure}

\begin{figure}
\centerline{\epsfig{file=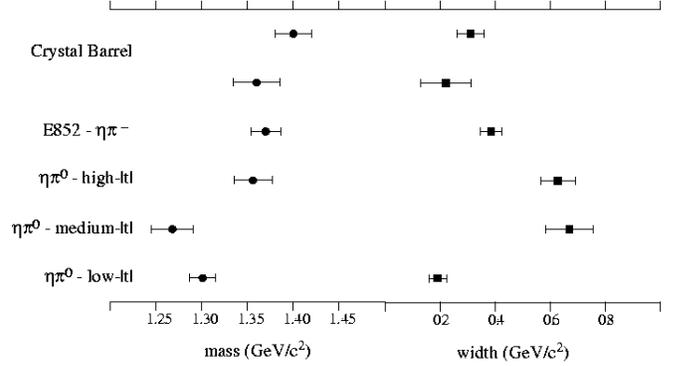,width=\columnwidth}}
\caption{Comparison of P-wave mass and widths among various studies.
A mass dependent fit to the $D_+$ and  $P_+$ amplitudes
and $\Delta \Phi(P_+ - D_+)$ phase difference, assuming Breit-Wigner
line shapes was carried out for the E852 data for $\eta \pi^0$ - this study - and
$\eta \pi^-$.  Results are also compared to the Crystal Barrel results.}
\label{pwavefits}
\end{figure}

The dependence  of the measured $D_+$ and $P_+$ amplitudes and the $\Delta \Phi(P_+ - D_+)$
phase difference as a function of $\eta \pi^0$ mass was fitted to
 two interfering Breit-Wigner
line shapes following the method of reference \cite{SUC1999}.
Fits carried out separately for the three ranges in $t$ are presented
in the plots of figure~\ref{massdep} and the fit parameters are listed in table~\ref{mdf25}.
 In figure~\ref{pwavefits} the $P$-wave parameters obtained from
the mass dependent fits are compared to those  reported in references \cite{SUC1999,Abe98,Abe99}.
 The $P$-wave mass obtained for the high-$t$ range is consistent
with that reported in the $\eta \pi^-$ channel but the width observed here is
significantly higher.  The values obtained for the mass in the low-$t$ and
high-$t$ ranges are lower than for the $\eta \pi^-$ channel.  
The conclusion is that no consistent set of resonant $P$-wave
parameters can describe the data while the resonant parameters obtained for the
$a_2(1320)$ are consistent for various ranges in $\eta \pi^0$ mass and $|t|$ and
consistent with well-established parameters from earlier experiments.

\begin{table}
\begin{tabular}{lccccc}

\hline \hline

            & all $|t|$    & low-$|t|$         & medium-$|t|$            & high-$|t|$    \\ \hline \hline

$M_{a_2}$     & 1.326  & 1.316          & 1.329          & 1.326   \\

            & $\pm 0.0023$    &$\pm 0.0049$    &$\pm 0.0029$     & $\pm 0.0036$  \\ \hline

$\Gamma_{a_2}$ & 0.169  &0.127           & 0.154         & 0.166& \\

            &$\pm 0.0069$    & $\pm 0.014$     &$\pm 0.0082$   &  $\pm 0.01$ \\ \hline

$M_{X}$      & 1.272   &1.301           & 1.268          & 1.356 \\

            & $\pm 0.017$    &$\pm 0.014$     & $\pm 0.023$   & $\pm 0.021$  \\ \hline

$\Gamma_{X}$ &0.66   & 0.19           &0.67          & 0.629 \\

             & $\pm 0.048$    &$\pm 0.032$     &$\pm 0.087$    &$\pm 0.064$  \\ \hline

$\chi^2_{\nu}$  & 3.23  &2.13        & 1.51        &  1.60   \\ \hline \hline

 \end{tabular}
 \caption{The masses and widths obtained from  a simultaneous fit to $D_{+}$ intensity, the $P_{+}$
intensity and the $\Delta \Phi(P_{+}-D_{+})$ phase difference. For these fits both waves are parametrized as
Breit-Wigner resonances. In addition to the fits for the three ranges of $t$ we also include the fit for
the full $t$ range to compare with the results of reference~\cite{SUC1999}.}
\label{mdf25}
\end{table}

\subsection{Description of the $t$ dependence of the  $D$-waves}

\begin{figure}
\centerline{\epsfig{file=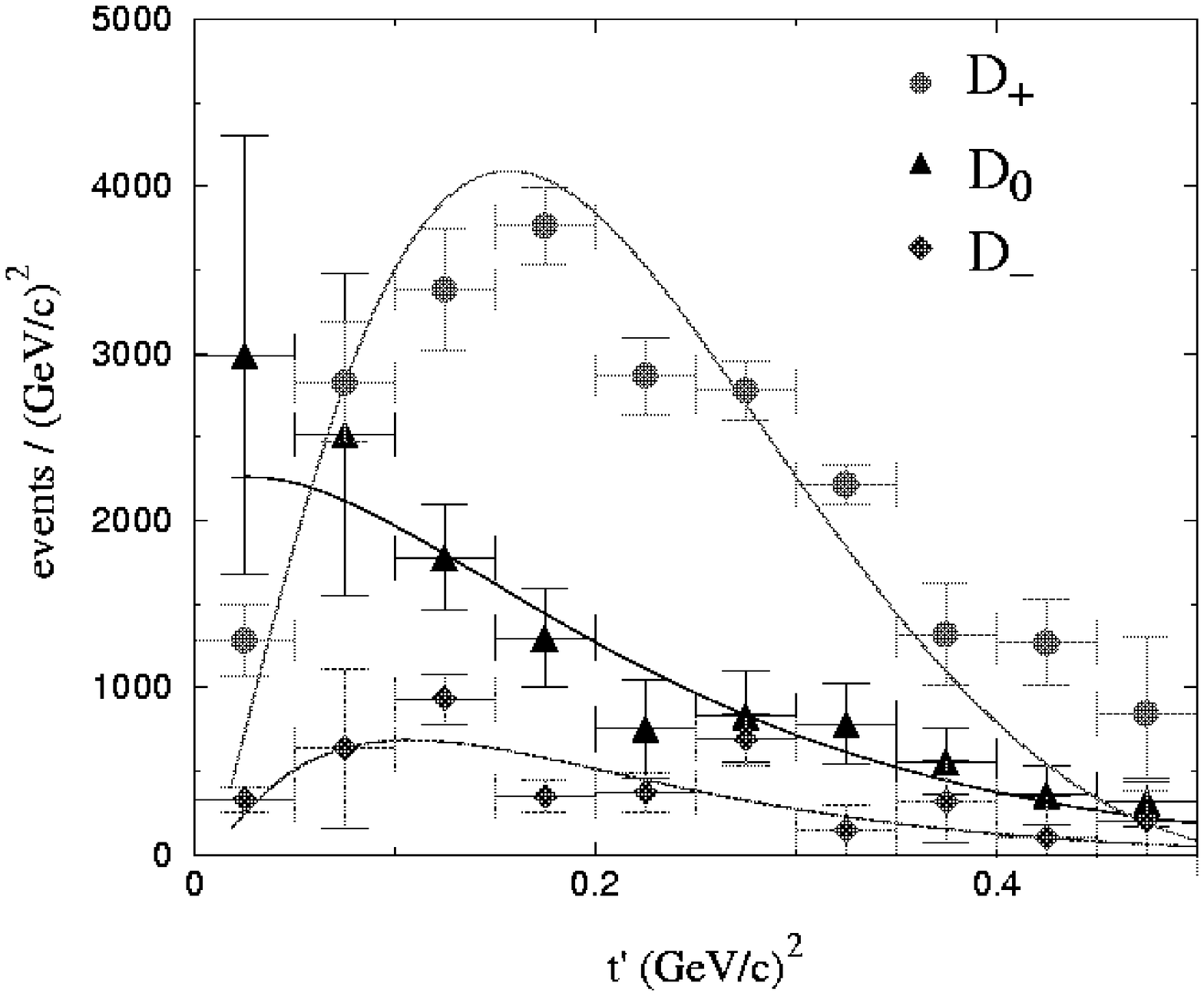,width=\columnwidth}}
\caption{Dependence of $D_+$ (circles), $D_0$ (triangles)
and $D_-$ (diamonds) waves  on  $|t^{\prime}|$
or $|t_{\pi^{-}\to\eta\pi}| - |t_{min}|$. The curves are the results of fits to forms
for these distributions expected from Regge phenomenology.}
\label{tdep}
\end{figure}

Figure \ref{tdep} shows the dependence of the $D_+$, $D_0$ and  $D_-$-waves as 
a function of $t_{\pi^{-}\to\eta\pi}$.  Since the $a_2(1320)$ is dominant
in these waves and the production of these waves is associated with exchanges
well known from Regge phenomenology, a comparison of data with this known
phenomenology is a check on the methodology.  Indeed the $t$ dependence is
well described.

The $D_+$ wave is produced via a natural $t$-channel charge exchange 
which is dominated by the $\rho$ trajectory and is parametrized by, \cite{IW}

\begin{eqnarray}
\nonumber D^{++}_+ =  i\sqrt{-{{t'}\over {4m_N^2}}}
  \left(1 - {t\over m_a^2}\right)^2
 e^{b_\rho t}e^{-i\pi{{\alpha_\rho(t)}/2}} \times \\
\Gamma(1 - \alpha_\rho(t)) \sin(\pi \alpha_\rho(t)/2)
\end{eqnarray}
\begin{equation}
D^{+-}_+ = r \sqrt{-t'\over {4 m_N^2}} D^{++}_+.  \label{d+}
\end{equation}

Here the $D^{+-}_+$ and $D^{++}_+$ are the nucleon helicity flip and
non-flip amplitudes respectively, with  $r=7$ in equation~\ref{d+}
according to reference~\cite{IW}, and 

\begin{equation}
|D_+|^2 = |D^{++}_+|^2 + |D^{+-}_+|^2. 
\end{equation}
The parameters of the $\rho$ trajectory are given by 
 $\alpha_\rho(t) = 0.5 + 0.9 t\, [\mbox{GeV}^2]$, $b_\rho = 3.25 \mbox{ GeV}^{-2}$ \cite{IM}.  
The fit of the  $t$-distribution given by Eq.~(\ref{d+}) to the 
 measured  $t$-distribution involves one parameter -- the overall
 normalization and is shown in figure ~\ref{tdep}. Some discrepancy is
 seen at large- $t$, in particular the characteristic dip of the $\rho$
 exchange at $t=m_\rho^2$ is not seen. This indicates that there is
 some absorption required for the $a_2$. 

The $D^0$ wave is produced via unnatural exchange and is expected to
 be dominated by the $b_1$ in the nucleon helicity flip
 amplitude  and the tensor, $Z$-trajectory, ($J^{PC} I =2^{--} 1$) 
 in the non-flip amplitude. These can parametrized by, \cite{IW}

\begin{eqnarray}
\nonumber D^{+-}_0 =  i  \sqrt{-t'\over {4 m_N^2}} 
  \left(1 - {t\over m_a^2}\right)
 e^{b_{b_1} t}\times
e^{-i\pi{{\alpha_{b_1}(t)}/2}} \times \\
\Gamma( - \alpha_{b_1}(t)) \sin(\pi \alpha_{b_1}(t)/2) \\
D^{++}_0 =  
 e^{b_{Z} t}e^{-i\pi{{\alpha_{Z}(t)}/2}} 
\Gamma(1 - \alpha_{Z}(t)) \cos(\pi \alpha_{Z}(t)/2).  \label{d0}
\end{eqnarray}
with the parameters of the two trajectories given by 
 $\alpha_{b_1} = -0.37 + 0.9t, [\mbox{GeV}^2]$,  $\alpha_{Z} = 0.9t,
 [\mbox{GeV}^2]$, and $b_{b_1} = b_Z = b_\pi = 4.51 \mbox{ GeV}^{-2}$ 
\cite{IM} .

The $D_-$-wave is produced by unnatural parity exchange and its
$t$ dependence is given by $D_-(t)=\sqrt{-t'/4m_N^2} \dot D_0(t)$.
The fit to the measured $D_0$ wave
 intensity has two parameters -- the normalization of the nucleon
 helicity flip and non-flip amplitudes -- and is also shown in figure ~\ref{tdep}.

\subsection{Ratio of $D$-waves}

In a paper discussing the PWA of the $\eta \pi$ system produced in peripheral
$\pi p$ interactions \cite{Sad99}, Sadovsky uses a criterion for selecting 
 the physical solution among the ambiguous solutions.  The test involves
a measurement of the ratio $r_S={{\left( {D_0-D_-} \right)} \mathord{\left/ 
{\vphantom {{\left( {D_0-D_-} \right)} {D_+}}} \right. \kern-\nulldelimiterspace} {D_+}}$.
According to Regge phenomenology this ratio should fall like 
$\alpha / p_{\pi}$
where $\alpha$ is determined from other experiments.  At $p_{\pi}$=18~GeV$/c$\ this
ratio is expected to be $r_S=0.80$.  Integrating the distributions in figure~\ref{tdep}
over all $t$ we obtain $r_S=0.72 \pm 0.12$.

\section {Conclusions}

 A partial wave and moment (averages of spherical
harmonics) analysis  of data collected in experiment
 E852 from the  reaction $ \pi^- p \rightarrow \eta \pi^0 n$
 (where $\eta \to \gamma \gamma$) as a function of $\eta \pi^0$
 effective mass and momentum-transfer-squared was performed.
 The presence of two well-established $a_0(980)$ and $a_2(1320)$
 resonances allows for the imposition of criteria for selecting the physical solution
 among the mathematically ambiguous solutions.
 The physical solution is consistent with the presence of two
 interfering Breit-Wigner resonances for these states.
The distribution in 12 acceptance-corrected moments, $H(L,M)$,  calculated
 directly from the data are consistent with moments calculated from
 the PWA solutions.

Every term in the moments with {\it odd} $L$ (5 of the 12) is linear in the $P$ wave
and the observed moments distributions are non-zero 
clearing indicating that the data demand the
presence of a $P$ wave.  To allow for the inclusion of a possible
resonance with exotic quantum numbers, the moments are thus fit with
the assumption of three interfering Breit-Wigner 
 resonances for the $S$-wave and $D$-wave and $P$-wave.

The data suggest weak production of the $a_0(1450)$. The $a_0(1450)$
was established by the Crystal Barrel collaboration  in $\bar p p \to
 \pi^0\pi^0\eta$ and $\bar p p \to \pi^0\eta\eta$ \cite{a01450} with production
ratios for
 $a_0(1450):a_0(980):a_2(1320)$ equal to  $4:6:33$
 and $11:14:25$ respectively.  These are qualitatively consistent with our
results.
 The dominant $D_+$, $D_0$ and $D_-$-waves are consistent with
 production of the $a_2(1320)$ and fits to the their line shapes are
 consistent with the resonance parameters of this well-established
 tensor resonance.  Moreover, the observed  $t$-dependence  of the
 individual $D$ waves are well described by Regge phenomenology as is
 their ratio. Finally the $a_0$ and $a_2$ in the data
 are well described as Breit-Wigner resonances produced via a
 $t$-channel exchange mechanisms regardless the treatment of the
 $P$-wave, and/or incoherent background.

 While the Breit-Wigner
 parameterization of the $S$ wave and $D$ wave is physically well
  motivated and confirmed by the data, the resonance representation
 of the $P$-wave is problematic. The data unambiguously
  indicated that the phase of the $P$-wave increases
 as the function of the $\eta\pi^0$ mass which corresponds to an
 attractive interaction in the $\eta\pi^0$ $P$-wave channel. However,
 the data do not require the phase of the $P$-wave
 to increase over $90^o$ nor  is there compelling evidence for the
   intensity of the $P$-wave to have a resonance shape in the
  mass range studied, {\it i.e} from threshold to $1.8\mbox{ GeV}$.

Mass-dependent fits, restricted to the $P_+$ and $D_+$ amplitudes and
$\Delta \Phi(P_+ - D_+)$
phase difference, following the method of reference~\cite{SUC1999}
reporting an exotic
$J^{PC}=1^{-+}$ meson decaying into $\eta \pi^-$, yield results
inconsistent with
those of  reference~\cite{SUC1999} and among the three $t$ ranges in
this analysis. The discrepancy in the resonance parameters
 found by fitting the natural waves alone is, however, smaller than
 the one found by fitting the full data sample (moments).

However, by restricting the study to only $P_+$, $D_+$ waves and the phase,
 $\Delta \Phi(P_+ - D_+)$, important additional information from the 
remaining data
is ignored.  For example, 
 an exotic $P$ wave resonance 
  should have the same resonance parameters in all three, $P_-$,$P_0$ and
 $P_+$-waves.
These wave represent different spin projections of the
 resonance and the data should discriminate between their production
 strengths. 

 Furthermore, we have found that
 the intensity of the weak waves, {\it i.e.} $P$ and $S$-wave above
 the  $a_0$ region are more  affected by leakage from the strong,
 $D$ waves due to the  $a_2$ than are the phase differences. Thus, since the
 mass-dependent fit to the  natural waves alone, is  strongly
 constrained by the intensity distribution of the $P$-wave, it tends to
 follow the $a_2$ line shape, and force the fit to
 mimic a $P$-wave resonance with mass near the $a_2$ mass.

In a separate paper we will present an alternative description of the
mass dependent
$P$-wave amplitude and phase that does not require the existence of
an exotic meson but is consistent
with $\eta \pi$ re-scattering.

\section {Acknowledgments}

The authors wish to thank the members of the MPS group at BNL as well as the staffs of the 
AGS and BNL.  This work was supported in part by the National Science Foundation
and the U. S. Department of Energy.

\pagebreak

\end{document}